%% file: main.tex
\documentclass[11pt]{article}

\usepackage[left=1in, top=1in, right=1in, bottom=1in, head=.1in, nofoot]{geometry}
\usepackage{setspace}
\usepackage{titlesec}
\titlelabel{\thetitle.\quad}
\usepackage{amsmath, amssymb, amsthm}
\usepackage{graphicx}
\graphicspath{{./}}
\usepackage{xcolor}
\usepackage[margin=20pt]{subcaption}
\usepackage{booktabs}
\usepackage{multirow}
\usepackage{placeins}
\usepackage{float}
\usepackage{microtype}

\usepackage{natbib}
\usepackage{xurl}
\usepackage{hyperref}
\hypersetup{
    colorlinks=true,
    linkcolor=black,
    citecolor=blue,
    urlcolor=blue
}
\usepackage[english]{babel}

\newtheorem{theorem}{Theorem}
\newtheorem{lemma}{Lemma}
\newtheorem{proposition}{Proposition}

\theoremstyle{definition}

\newtheorem{assumption}{Assumption}
\newtheorem{remark}{Remark}

\numberwithin{equation}{section}
\numberwithin{figure}{section}
\numberwithin{table}{section}
\numberwithin{theorem}{section}
\numberwithin{lemma}{section}
\numberwithin{proposition}{section}
\numberwithin{corollary}{section}
\numberwithin{definition}{section}
\numberwithin{assumption}{section}
\numberwithin{remark}{section}

\newcommand{\E}{\mathbb{E}}
\newcommand{\Prob}{\mathbb{P}}

\newcommand{\var}{\text{Var}}

\newcommand{\indep}{\perp \!\!\! \perp}

\title{\bfseries Covariate-Adaptive Enrollment for Single-Arm Trials with External Controls}
\author{
\small
Jiahui Xin$^{1}$, \ \ Hengjia Fang$^{2}$, \ \ Zhiyue Huang$^{3}$, \ \
Satoshi Yoshida$^{4}$, \ \ Wei Ma$^{2}$%
\thanks{\small Corresponding author: \texttt{mawei@ruc.edu.cn}}
\\[1ex]
{\small $^{1}$ The University of Texas MD Anderson Cancer Center, Houston, Texas, USA}\\
{\small $^{2}$ Institute of Statistics and Big Data, Renmin University of China, Beijing, China}\\
{\small $^{3}$ Quantitative Science and Evidence Generation, Astellas Pharma Inc., Shanghai, China}\\
{\small $^{4}$ Quantitative Science and Evidence Generation, Astellas Pharma Inc., Tokyo, Japan}
}

\begin{document}
\doublespacing

\maketitle

\begin{abstract}
Single-arm trials with external controls now support regulatory decisions, but the enrolled patients may differ in baseline covariates from the external population. Retrospective weighting analysis after enrollment can align measured covariate summaries, but possible extreme weights induced by covariate shift can reduce effective sample size and statistical power, which makes the trial hard to plan. We propose covariate-adaptive enrollment based on a biased-coin rule, a prospective design that uses each candidate's baseline covariates to favor candidates whose enrollment brings the cohort's covariate summaries closer to the external targets. We establish conditions for strong balance in the targeted covariate summaries, thereby addressing measured covariate shift by design. For the one-sample log-rank test under this adaptive design, we derive the asymptotic null distribution and a consistent variance estimator, and show that the conventional variance estimator is conservative under strong balance. Eflornithine, the first oncology drug approved by the FDA primarily on evidence from an externally controlled trial, motivates our hypothetical design evaluation in high-risk neuroblastoma. In this evaluation based on published evidence, adaptive enrollment gave a higher effective sample size and higher power than ordinary enrollment followed by matching-adjusted indirect comparison. Further simulations with time-to-event and continuous endpoints examine covariate shift and unmeasured confounding.

\vspace{12pt}
\end{abstract}

\section{Introduction}

Regulatory authorities and health technology assessment (HTA) bodies increasingly evaluate evidence from single-arm trials (SATs) when randomized comparisons are infeasible or unethical \citep{ema2024satreflection,fda2023ectguidance,mishrakalyani2022external,patel2021hta}. Between 2019 and 2022, submissions supporting 37 non-oncology approvals, 20 by the U.S. Food and Drug Administration (FDA) and 17 by the European Medicines Agency, contained a pivotal SAT \citep{Subramaniam2024}. An SAT alone gives no relative treatment effect, so HTA submissions based on single-arm evidence use external controls (ECs) as the comparator \citep{patel2021hta}. The credibility of the resulting indirect comparison depends on how comparable the trial cohort is with the EC population.

High-risk neuroblastoma is a rare, life-threatening pediatric malignancy. In 2023 the FDA approved eflornithine (DFMO) for certain adult and pediatric patients with high-risk neuroblastoma \citep{fda2023eflornithineapproval}, the first oncology drug approval to rely on an externally controlled trial as the primary clinical evidence of effectiveness \citep{duke2024eflornithine}. The single-arm Study 3(b) was planned to compare two-year event-free survival (EFS) with a historical rate of 70\% from the ANBL0032 trial \citep{fda2023briefing}. U.S. regulations require an externally controlled analysis to be based on comparable patients or populations, and in 2018 the FDA recommended a comparison with patient-level ANBL0032 data \citep{fda2023briefing}. The approval analysis matched patients on the propensity score and MYCN status: of 105 eligible patients treated in Study 3(b), 90 were matched to 270 of the 1,241 ANBL0032 patients. Published EC evidence can provide two planning inputs even when original patient records are unavailable. Aggregate summaries describe selected baseline features of the EC population, and Kaplan--Meier curves support an approximate survival benchmark. The reported moments do not identify the joint EC covariate distribution, and the composition of the future trial cohort becomes known only during screening. Differences in the reported covariate summaries can be reduced during enrollment rather than addressed only at the analysis stage.

Matching-adjusted indirect comparison (MAIC) is widely used when only aggregate comparator data are available. Of 32 NICE oncology appraisals from 2017 to 2022 that relied only on SATs, 21 used MAIC to produce comparative evidence and 12 used a naive comparison \citep{zou2022use}. MAIC weights individual trial records so that selected covariate moments agree with published EC summaries \citep{signorovitch2012matching,cheng2020maic,cheng2023maic}. MAIC acts after the data are collected. When overlap is limited, the weights become extreme \citep{lambert2023enriching}, and both the effective sample size (ESS; \citealp{kish1965survey}) and the precision of MAIC decrease \citep{remiroazocar2022twostage}. The single-arm CARTITUDE-1 trial of the CAR T-cell therapy ciltacabtagene autoleucel illustrates the size of this loss. When its 97 treated patients were weighted to the DREAMM-2 population, the ESS fell to 39, a 60\% reduction \citep{weisel2022matching}.

This loss is costly in rare diseases. Treatment and longitudinal monitoring in high-risk neuroblastoma are resource intensive \citep{shoag2024utilization}, so information discarded by weighting was collected at substantial expense. Because the overlap is unknown before accrual, the weighted ESS, the nominal enrollment and the power are also hard to plan. Sample size methods for single-arm survival trials test the trial against a fixed reference survival distribution, which must be correctly specified, and make no adjustment for covariate differences between the trial and reference populations \citep{wu2014sample,dai2024sample}. The question we address is whether reported EC summaries can guide enrollment so that the cohort is comparable on the measured features and little information is lost to weighting.

We propose covariate-adaptive enrollment (CAE) for this purpose. During screening, CAE compares the current imbalance with the imbalance that enrolling an eligible candidate would produce, and a biased coin favors candidates who move the cohort toward the reported EC moments. Enrollment decisions use only prespecified baseline features measured before treatment is administered and before outcomes are observed. Final weighting then calibrates the analysis to the external target. By bringing the cohort closer to the external target, CAE aims to reduce weight variability and preserve ESS. The rule builds on covariate-adaptive randomization \citep{efron1971forcing,pocock1975sequential,wei1978adaptive,hu2012asymptotic,ma2024new}, but it aligns one enrolling cohort with an external target instead of balancing randomized arms. Because CAE changes how the cohort is formed, the analysis must account for the enrollment rule.

This article makes three contributions. First, we propose a biased-coin enrollment rule based on aggregate EC covariate summaries. Under feasibility and drift conditions, we show that its cumulative imbalance remains stochastically bounded, establishing strong balance regardless of endpoint type. Second, we establish asymptotic inference for survival and continuous outcomes. For the weighted one-sample log-rank score, we derive the asymptotic null distribution and a consistent variance estimator, and characterize when the conventional variance estimator is conservative. Third, we evaluate CAE as the design of a future high-risk neuroblastoma trial, with published DFMO summaries by prior treatment as the target and records reconstructed from Kaplan--Meier curves as the pooled survival benchmark. The evaluation compares CAE with ordinary enrollment, both followed by MAIC weighting. Further simulations with survival and continuous outcomes examine covariate shifts and the consequences of unmeasured confounding.

Section~\ref{sec:case-study} presents the neuroblastoma design, its published inputs, operational pipeline and primary test. Section~\ref{sec:methodology} gives the enrollment rule and the survival analysis. Section~\ref{sec:case-operating} evaluates the design by simulation, and Section~\ref{sec:theory} gives the theoretical results. Section~\ref{sec:numerical} examines additional population shifts. Section~\ref{sec:discussion} discusses practical implications, limitations and future implementation. Simulation details, supporting studies, sensitivity analyses, regularity conditions and proofs are in the appendix.

\section{A Prospective High-Risk Neuroblastoma Trial Design Using Published DFMO Evidence\label{sec:case-study}}

We consider a future single-arm trial of an investigational maintenance treatment for high-risk neuroblastoma, with no concurrent control group. The external information used for enrollment and primary analysis consists of a pooled survival benchmark and aggregate covariate summaries. The benchmark describes survival in the population represented by those summaries. If the enrolled population has a different composition, its pooled survival may differ from the benchmark even when survival within each prognostic group is unchanged. The design question is therefore how to use the external summaries during enrollment to form a cohort suitable for this comparison.

Published DFMO evidence provides a clinical setting for this question. In 2023, eflornithine (DFMO) became the first oncology drug approved by the FDA with an externally controlled trial providing the primary clinical evidence of effectiveness \citep{duke2024eflornithine}. Although Study 3(b) was prospectively designed to compare event-free survival (EFS) with a historical benchmark, population comparability remained central to its regulatory review. The FDA noted that ``Comparability of the external control arm to the investigational arm was a key review issue'' \citep[p.~31]{fda2023briefing}.

We use the published DFMO cohorts as the reference population for the future trial. The intention-to-treat population comprised 100 patients after standard high-risk therapy (Stratum 1) and 39 after treatment for relapsed disease or augmented treatment for primary refractory disease (Stratum 2) \citep{sholler2018dfmo}. Here the covariate is the stratum, so the aggregate summary is the vector of stratum proportions, $p^*=(100/139,\,39/139)\approx(0.72,\,0.28)$. It defines the target population and serves as the enrollment target $\boldsymbol{\mu}_{EC}$. As in the DFMO publication, EFS is measured from the first dose of maintenance treatment \citep{sholler2018dfmo,hernan2016target}. The estimand is the EFS distribution of the investigational treatment in the target population. The null hypothesis is that it equals the survival benchmark $S_0=\exp(-\widehat{\Lambda}_{EC})$ described below. With a total enrollment of 100 patients, we compare two enrollment rules, ordinary enrollment and CAE. The ordinarily enrolled cohort is analyzed using the unadjusted one-sample log-rank test (OSLRT) and the MAIC-weighted OSLRT. The CAE cohort is analyzed as described below.

\begin{figure}[!t]
\centering
\includegraphics[width=0.99\linewidth]{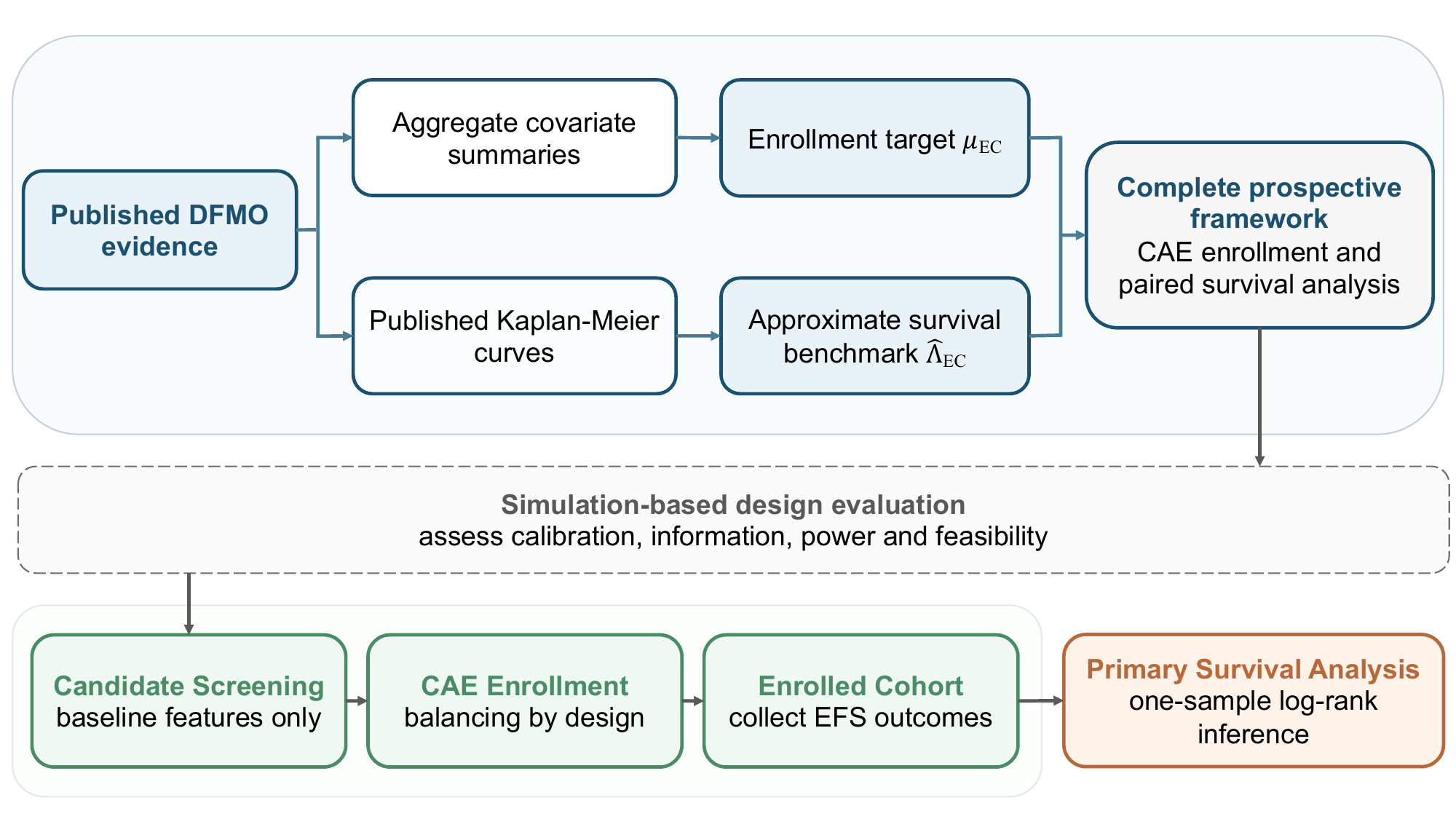}
\caption{Workflow of the proposed neuroblastoma design. The published DFMO evidence enters in two ways: the aggregate summaries set the enrollment target $\boldsymbol{\mu}_{EC}$, and the published Kaplan--Meier curves give the comparison benchmark $\widehat{\Lambda}_{EC}$. The design is first evaluated by simulation. The future single-arm trial would then screen candidates, enroll them by CAE, collect EFS, and test against the comparison benchmark.}
\label{fig:cae-workflow}
\end{figure}

Individual DFMO records are not available to us. We instead use approximate event and censoring records reconstructed from the published Kaplan--Meier curves, which we call pseudo individual patient data (pseudo-IPD) \citep{guyot2012ipd,liu2021ipdfromkm}. We pool these records, fit a spline survival curve, and use its cumulative hazard $\widehat{\Lambda}_{EC}$ as the fixed comparison benchmark. The stratum-specific EFS values are also used to build the simulation model; they are not used in the primary analysis. Figure~\ref{fig:cae-workflow} summarizes the workflow.

In the evaluation, ordinary enrollment and CAE share a candidate arrival process with Poisson rate 50 per year, and each candidate is equally likely to come from either stratum. The candidate population therefore differs from the external population, which has about 72\% of patients in Stratum 1. Ordinary enrollment accepts the first 100 candidates. CAE enrolls every candidate until both strata have appeared. It then uses a biased-coin rule similar to that of \citet{efron1971forcing}. A candidate is enrolled with probability 1.00 if enrolling the candidate reduces the squared cumulative imbalance from $\boldsymbol{\mu}_{EC}$ defined in Section~\ref{sec:methodology}, 0.75 if it leaves the imbalance unchanged, and 0.50 if it increases the imbalance. Screening continues until 100 patients are enrolled. Under either rule, the cohort is analyzed two years after its last patient is enrolled.

\begin{figure}[!t]
\centering
\includegraphics[width=\linewidth]{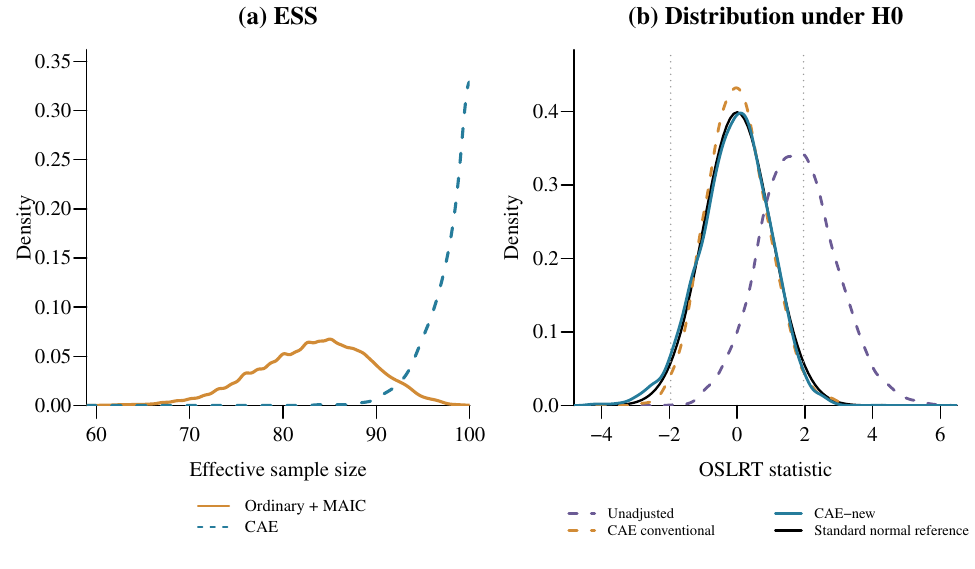}
\caption{First look at the DFMO evaluation. Left: effective sample size (ESS) under ordinary enrollment followed by MAIC and under CAE. Right: null distributions of the unadjusted OSLRT, CAE-conventional and CAE-new. The black curve is the standard normal density, and the vertical lines mark $\pm1.96$. CAE-conventional and CAE-new use the same cohort and the same score and differ only in the variance.}
\label{fig:dfmo-info}
\end{figure}

Because the CAE cohort is already close to the target, it needs less adjustment after enrollment. The cohort is calibrated to $p^*$, and a weighted OSLRT compares it with the benchmark. Our primary analysis, CAE-new, standardizes the weighted score using the design-aware variance estimator and tests at a two-sided nominal level of 0.05. Two variants use the same CAE cohort. CAE-conventional replaces the design-aware variance with the conventional one, which shows the effect of the variance alone. CAE-JK uses jackknife inference for the conventional OSLRT statistic; it serves as a sensitivity analysis and provides a comparison with ordinary enrollment followed by MAIC and jackknife inference (MAIC-JK). Figure~\ref{fig:dfmo-info} gives an initial picture of the design. It shows the effective sample size (ESS) and the null distributions of the unadjusted OSLRT, CAE-conventional and CAE-new. Section~\ref{sec:case-operating} further compares ESS and rejection probabilities between CAE and ordinary enrollment followed by MAIC.

\section{Covariate-Adaptive Enrollment and Subsequent Analysis \label{sec:methodology}}

CAE is a prospective enrollment design that uses aggregate external covariate summaries to guide the composition of the internal trial cohort. Because enrollment decisions depend on baseline covariates and previous enrollment decisions but not on outcomes, the same rule can be used with different endpoint types. Categorical and continuous covariates enter through a prespecified feature vector, whose external target summaries define the balance objective.

Motivated by the DFMO application, we develop a weighted one-sample log-rank analysis for finitely many strata. Inference and simulations for continuous endpoints are presented in Appendix~\ref{sec:supp-continuous}.

\subsection{Enrollment Using Aggregate EC Covariate Summaries}

The design is defined over the sequence of screened candidates. For the balance theory, their baseline covariates are viewed as an i.i.d.\ sequence from a fixed superpopulation, $X_i$, $i=1,2,\ldots$. The baseline covariate $X_i$ is observed before the enrollment decision, whereas outcomes are observed only for enrolled SAT subjects and are introduced in the endpoint-specific sections. The design covariate $\boldsymbol{x}_i$ denotes the fixed-dimensional feature vector $b(X_i)$ selected for prospective balance between SAT and EC. Future candidate covariates are independent of the past enrollment history. Conditional on the past history and on the currently observed covariate, the enrollment indicator is generated according to the biased-coin probability below.

Let $N$ denote the total number screened in the enrollment. Patients are screened sequentially with covariate vector $\boldsymbol{x}_n$. The design input $\boldsymbol{\mu}_{EC}$ is the reported EC target vector constructed from prespecified aggregate summaries. Means, categorical proportions, and higher-order moments can all be represented by choosing the feature map $b(X)$ appropriately. The cumulative imbalance at the $n$-th patient is defined as
\begin{equation}
    \boldsymbol{D}_n
    = \sum_{i=1}^n Z_i
      (\boldsymbol{x}_i-\boldsymbol{\mu}_{EC}),
\end{equation}
where $Z_i\in\{0,1\}$ indicates enrollment ($Z_i=1$ implies the patient is enrolled into the SAT). Let $m_n=\sum_{i=1}^n Z_i$ be the number enrolled after screening $n$ candidates, and write $m=m_N$ for the final enrolled sample size. Whenever $m_n>0$, the enrolled covariate mean is
\[
    \bar{\boldsymbol{x}}_{SAT}^{(n)}=m_n^{-1}\sum_{i=1}^n Z_i\boldsymbol{x}_i,
    \qquad
    \boldsymbol{D}_n=m_n\bigl(\bar{\boldsymbol{x}}_{SAT}^{(n)}-\boldsymbol{\mu}_{EC}\bigr).
\]

Let $\mathcal{F}_{n-1}=\sigma(\boldsymbol{x}_1,Z_1,\dots,\boldsymbol{x}_{n-1},Z_{n-1})$ be the history of covariates and enrollment decisions for the first $n-1$ candidates. When candidate $n$ arrives with covariate $\boldsymbol{x}_n$, the design compares the imbalance with and without this candidate. Enrolling the candidate changes the squared Euclidean norm of the imbalance by
\[
  \delta_n
  =\|\boldsymbol{D}_{n-1}+\boldsymbol{x}_n-\boldsymbol{\mu}_{EC}\|^2-\|\boldsymbol{D}_{n-1}\|^2
  =2\boldsymbol{D}_{n-1}^{\top}(\boldsymbol{x}_n-\boldsymbol{\mu}_{EC})
   +\|\boldsymbol{x}_n-\boldsymbol{\mu}_{EC}\|^2 .
\]
The candidate is enrolled with probability
\begin{equation}
  \label{eq:design_function}
  \phi_n=\Prob(Z_n=1\mid\mathcal{F}_{n-1},\boldsymbol{x}_n)
  =p_{\mathrm{avg}}-\frac{\Delta p}{2}\operatorname{sgn}(\delta_n)
  =\begin{cases}
     p_{\mathrm{avg}}+\Delta p/2, & \delta_n<0,\\
     p_{\mathrm{avg}},            & \delta_n=0,\\
     p_{\mathrm{avg}}-\Delta p/2, & \delta_n>0,
   \end{cases}
\end{equation}
where $0<\Delta p/2 < p_{\mathrm{avg}}<1$ and $p_{\mathrm{avg}}+\Delta p/2\le1$, so that all three probabilities lie in $(0,1]$. Thus a candidate who moves the cohort toward $\boldsymbol{\mu}_{EC}$ is more likely to be enrolled than one who moves it away. This is the biased-coin idea of \citet{efron1971forcing}, applied to the decision to enroll rather than to treatment assignment. The parameter $p_{\mathrm{avg}}$ sets the baseline acceptance probability, and $\Delta p$ sets how strongly the rule favors balance. The case study uses $p_{\mathrm{avg}}=0.75$ and $\Delta p=0.5$, giving acceptance probabilities of 1.00, 0.75 and 0.50, respectively.

\subsection{Survival Analysis after CAE\label{sec:inference}}

CAE introduces dependence among the enrolled SAT patients, so they are no longer an i.i.d.\ sample. Also, the biased-coin enrollment may not completely remove the covariate shift. We use the following calibration to address the remaining differences, and the variance estimation must reflect the resulting sampling structure. We therefore propose a one-sample log-rank test with a design-aware variance for finitely many strata. Section~\ref{sec:theory} establishes conditions under which the design-aware variance estimator is consistent and the conventional variance estimator is conservative. Corresponding inference and variance comparisons for continuous endpoints are given in Appendix~\ref{sec:supp-continuous}. In this section, we focus on the time-to-event endpoint, motivated by the DFMO case study.

The analysis compares the calibrated SAT cohort with the fixed EC survival benchmark. Let $m$ be the enrolled sample size, $s_i\in\{1,\ldots,J\}$ the stratum of subject $i$, $n_j$ the enrolled count in stratum $j$, and $p_j^*=p_{EC,j}$ its target proportion. Let $Y_i$ be the observed follow-up time and $\Delta_i$ the event indicator. The individual score contribution
\begin{equation}
    M_i=\Delta_i-\widehat\Lambda_{EC}(Y_i)
    \label{eq:individual-discrepancy}
\end{equation}
is the observed event indicator minus the benchmark cumulative hazard at $Y_i$ \citep{breslow1975analysis,fleming2011counting}. Write $\mu_j=\E(M_i\mid s_i=j)$ for the mean score contribution in stratum $j$. The score is centered when $\sum_jp_j^*\mu_j=0$. Under the conditions in Section~\ref{sec:theory}, the null hypothesis in Equation~\eqref{eq:survival-mixture-null} implies this centering condition.

After CAE, calibration gives each subject in stratum $j$ the weight $w_j=mp_j^*/n_j$, so the weights sum to $m$ and stratum $j$ receives total weight $mp_j^*$. When the only covariate is the stratum, these weights are the MAIC weights that match the EC proportions. The closer the enrolled proportions are to the target, the closer these weights are to one.

Let $\bar M_j$ and $s_j^2$ be the sample mean and variance of $M_i$ in stratum $j$, with divisor $n_j-1$. The weighted score, the design-aware variance and the proposed statistic are
\begin{equation}
    U=\sum_iw_iM_i=m\sum_jp_j^*\bar M_j,
    \qquad
    \widehat V_{\mathrm{new}}=\sum_jn_jw_j^2s_j^2=m^2\sum_j\frac{(p_j^*)^2s_j^2}{n_j},
    \qquad
    T_{\mathrm{new}}=\frac{U}{\sqrt{\widehat V_{\mathrm{new}}}}.
    \label{eq:cae-survival-statistic}
\end{equation}
Calibration fixes the total weight of each stratum at $mp_j^*$. Under the null, the stratum means $\mu_j$ cancel when weighted by $p_j^*$, so $U$ combines fluctuations of the sample means around their stratum means. The variance estimator $\widehat V_{\mathrm{new}}$ accounts for these fluctuations. The conventional weighted OSLRT uses the same score but estimates the variance by
\begin{equation}
    \widehat V_{\mathrm{conv}}
    =\sum_iw_i^2\widehat\Lambda_{EC}(Y_i),
    \qquad
    T_{\mathrm{conv}}=\frac{U}{\sqrt{\widehat V_{\mathrm{conv}}}}.
    \label{eq:survival-naive-var}
\end{equation}
We call the analysis based on $T_{\mathrm{new}}$ CAE-new, and the one based on $T_{\mathrm{conv}}$ in the same cohort CAE-conventional; comparing them shows the effect of the variance alone. For CAE-JK, we re-estimate the weights and recompute $T_{\mathrm{conv}}$ after deleting one patient at a time. We divide the full-sample $T_{\mathrm{conv}}$ by the square root of its jackknife variance, computed around the mean of the leave-one-out statistics. It serves as a sensitivity analysis and as the comparator for MAIC-JK, which applies the same jackknife to the MAIC-weighted OSLRT under ordinary enrollment. Section~\ref{sec:theory} gives the finite-stratum conditions and the interpretation of these variances.

\section{Simulation-Based Evaluation of the Proposed Neuroblastoma Trial Design\label{sec:case-operating}}

\begin{figure}[H]
\centering
\includegraphics[width=0.75\linewidth]{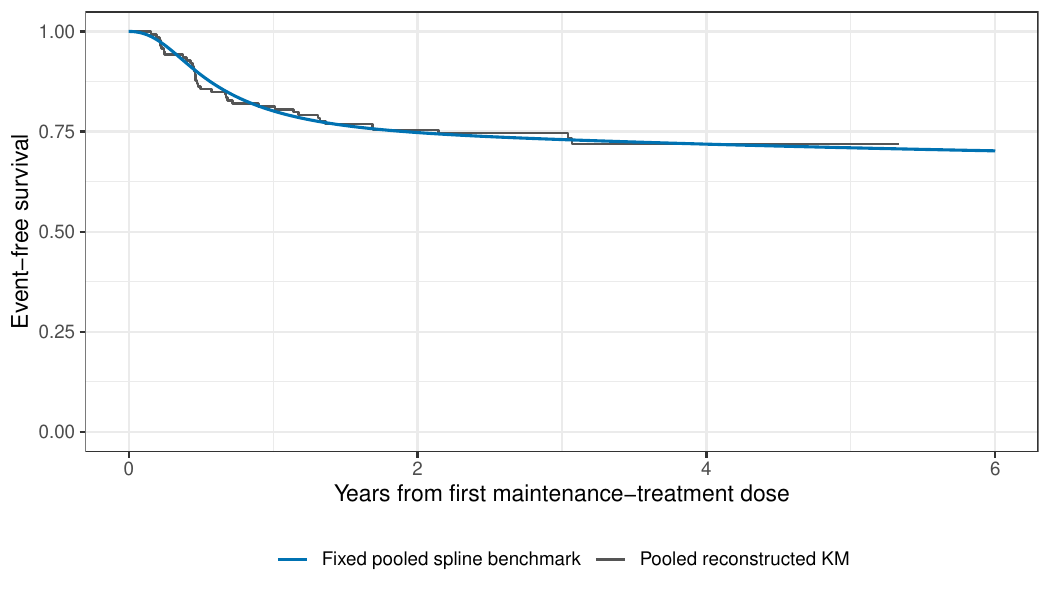}
\caption{External control for the DFMO evaluation. The step curve is the Kaplan--Meier estimate from the pooled DFMO pseudo-IPD. The smooth curve is the fixed spline survival benchmark $\exp(-\widehat\Lambda_{EC})$ with one internal knot.}
\label{fig:dfmo-ec-sat-km}
\end{figure}

This section evaluates the proposed design of Section~\ref{sec:case-study} with 5,000 replicates under each hypothesis. Figure~\ref{fig:dfmo-ec-sat-km} shows the external control. Under $H_0$, the data generating mechanism matches the reported two-year EFS of 0.84 in Stratum 1 and 0.51 in Stratum 2. Under $H_1$, the hazards are multiplied by 0.60 in Stratum 1 and 0.40 in Stratum 2 relative to their values under $H_0$. The benchmark and the target proportions are fixed across replicates. Appendix~\ref{sec:supp-survival-sim} gives the construction and the other survival analyses.

Table~\ref{tab:dfmo-oc} compares ordinary enrollment and CAE at the same enrolled sample size. Without adjustment, ordinary enrollment does not account for the covariate shift and has a high rejection probability under $H_0$. CAE gives a higher ESS and a higher $H_1$ rejection probability than ordinary enrollment followed by MAIC-JK in this evaluation.

Within the CAE cohort, conventional studentization is conservative in this evaluation. Design-aware studentization brings the standard deviation of the statistic under $H_0$ closer to one and increases $H_1$ rejection relative to conventional studentization. The jackknife gives similar results. These comparisons use the same patients and observed outcomes. Figure~\ref{fig:dfmo-info} shows the location shift of the unadjusted distribution and the difference in scale between the two CAE analyses.

\begin{table}[htbp]
\centering
\caption{DFMO evaluation with 100 enrolled patients. All tests are two-sided at the nominal level of 0.05. ESS is averaged across replicates.}
\label{tab:dfmo-oc}
\input{case_current_table.tex}
\end{table}

Section~\ref{sec:theory} gives the conditions for balance of the enrolled cohort and valid inference after calibration.

\section{Theoretical Guarantees\label{sec:theory}}

We establish balance properties of the enrollment rule and asymptotic properties of the calibrated survival analysis in Section~\ref{sec:inference}. A balance result gives conditions under which the enrolled cohort approaches the target $\boldsymbol{\mu}_{EC}$. A variance comparison explains when conventional studentization gives a conservative test. Throughout this section, the external proportions and the cumulative hazard are fixed.

\subsection{Design Balance between SAT and EC}

\begin{assumption}[Covariate Structure]
\label{ass:covariates}
The target mean $\boldsymbol{\mu}_{EC}$ lies strictly in the interior of the convex hull of the support of $\boldsymbol{x}$.
\end{assumption}

The candidate feature distribution $\Gamma$ satisfies either of the following two conditions.

\begin{assumption}[Continuous Covariate Distribution]
\label{ass:continuous}
The covariate vector $\boldsymbol{x}\in\mathbb R^k$ has finite second moment and a density $p\in L^2(\mathbb R^k)$ that is strictly positive on a connected open set containing $\boldsymbol\mu_{EC}$.
\end{assumption}

\begin{assumption}[Discrete Covariate Distribution]
\label{ass:discrete}
The covariate $\boldsymbol{x}$ takes finitely many distinct values $\boldsymbol{t}_1,\dots,\boldsymbol{t}_s$ with strictly positive probabilities, where $\{\boldsymbol{t}_i\}\subset\mathbb Q^k$ and $\boldsymbol{\mu}_{EC}\in\mathbb Q^k$.
\end{assumption}

Write $\boldsymbol{u}=\boldsymbol{x}-\boldsymbol{\mu}_{EC}$, $\boldsymbol{\delta}=\E(\boldsymbol{u})$, and
\[
    c_*=\inf_{\|\boldsymbol{v}\|=1}
    \E|\boldsymbol{v}^\top\boldsymbol{u}|.
\]
We require the feasibility condition
$$0<\Delta p<2\min\{p_{\mathrm{avg}},1-p_{\mathrm{avg}}\}$$
and the drift condition
\[
    \Delta p>
    \frac{2p_{\mathrm{avg}}\|\boldsymbol{\delta}\|}{c_*}.
\]
These conditions ensure stability of the imbalance process; Appendix~\ref{app:ergodicity} gives the Markov-chain argument.

\begin{theorem}[Asymptotic Balance between SAT and EC]
\label{thm:balance}
Under the CAE biased-coin design, Assumption~\ref{ass:covariates}, either Assumption~\ref{ass:continuous} or Assumption~\ref{ass:discrete}, and the feasibility and drift conditions above, the cumulative imbalance satisfies:
\begin{enumerate}
    \item[(i)]
$\|\boldsymbol{D}_N\|=O_p(1)$ and consequently $\bar{\boldsymbol{x}}_{SAT}^{(N)} -\boldsymbol{\mu}_{EC}=O_p(N^{-1})$.
    \item[(ii)]
If, in addition, the covariates are bounded, then $\|\boldsymbol{D}_N\|=O(\log N)$ a.s.\ and $\bar{\boldsymbol{x}}_{SAT}^{(N)} -\boldsymbol{\mu}_{EC}=O(N^{-1}\log N)$ a.s.
\end{enumerate}
\end{theorem}

The theorem establishes strong balance in the prespecified covariate features, whatever the endpoint type. Calibrated inference below requires positive limiting stratum proportions, but not design balance.

\subsection{A new one-sample log-rank test under CAE}

We use the strata, target proportions, score contributions and statistics defined in Section~\ref{sec:inference}, and write $\Lambda_0=\widehat\Lambda_{EC}$ and $S_0=\exp(-\Lambda_0)$. On a fixed follow-up interval $[0,\tau]$, assume that $S_0$ is positive and absolutely continuous, with $S_0(0)=1$ and $\Lambda_0(\tau)<\infty$. The number of strata $J$ and the positive target proportions $p_j^*$ are fixed.

For candidate $k$, let $s_k$ be the stratum, $F_k$ the potential event time and $C_k\le\tau$ the potential censoring time. The records $(s_k,F_k,C_k)$ are i.i.d., with a distribution that does not depend on $m$. Enrollment decisions depend only on candidate strata, previous enrollment decisions, and i.i.d.\ random variables $\xi_k\sim\mathrm{Unif}(0,1)$ that are independent of all candidate records. These variables implement the biased-coin rule: candidate $k$ is enrolled when $\xi_k$ is below the prescribed acceptance probability. Recruitment stops when $m$ patients have enrolled. The positive lower bound on the enrollment probability ensures that this occurs after finitely many candidates almost surely. The observed data are
\[
  Y_k=\min(F_k,C_k),\qquad
  \Delta_k=1\{F_k\le C_k\}.
\]
The score contribution $M_k$ is defined by Equation~\eqref{eq:individual-discrepancy}, with $i$ replaced by $k$. Write
\[
  S_j(t)=\Prob(F_k>t\mid s_k=j),\qquad
  \mu_j=\E(M_k\mid s_k=j),\qquad
  v_j=\var(M_k\mid s_k=j).
\]

\begin{assumption}[Censoring and Limiting Stratum Proportions]
\label{ass:finite-observation}
\leavevmode
\begin{enumerate}
\item[(a)]
The potential censoring time satisfies $C_k \indep (F_k,s_k)$. Each stratum-specific survival function $S_j$ is absolutely continuous on $[0,\tau]$.

\item[(b)]
Every candidate stratum has positive probability, $n_j/m\to_p q_j>0$ for deterministic $q_j$, and $v_j>0$ for every $j$.
\end{enumerate}
\end{assumption}

\begin{remark}
The censoring requirements in Assumption~\ref{ass:finite-observation}(a) are compatible with administrative censoring at a prespecified calendar date. If a participant enters at date $E$ and observation ends at date $a$, the potential censoring time is $C=a-E$. When entry dates have the same distribution across strata and are independent of event times within strata,
\[
  \Prob(C\ge t\mid s=j)
  =\Prob(E\le a-t\mid s=j)
  =G(t).
\]
Assumption~\ref{ass:finite-observation}(b) ensures that every stratum has a growing sample with the limiting enrolled proportions $q_j$ that may differ from the external targets $p_j^*$.
\end{remark}

We compare survival in an internal population with the external stratum proportions against the pooled EC benchmark. The null hypothesis is
\begin{equation}
  H_0:\quad
  \sum_{j=1}^Jp_j^*S_j(t)=S_0(t),
  \qquad 0\le t\le\tau.
  \label{eq:survival-mixture-null}
\end{equation}
Thus, if the internal population had the same stratum proportions as the EC population, its survival function, and hence its cumulative hazard, would equal the external benchmark under $H_0$.

With $G_j(t)=\Prob(C_k\ge t\mid s_k=j)$, independent censoring within strata gives
\[
  \mu_j=\E(M_k\mid s_k=j)
  =\int_0^\tau S_j(t)G_j(t)
    \{\lambda_j(t)-\lambda_0(t)\}\,dt.
\]
Here $\lambda_j(t)=-S_j'(t)/S_j(t)$ where $S_j(t)>0$ and zero otherwise. Under common $G$, Equation~\eqref{eq:survival-mixture-null} implies $\sum_jp_j^*\mu_j=0$ (Appendix~\ref{app:tte-regularity}).

The score uses the external proportions rather than the observed enrollment proportions. Under $H_0$,
\[
  U=m\sum_jp_j^*\bar M_j
   =m\sum_jp_j^*(\bar M_j-\mu_j).
\]
Thus $U$ combines fluctuations of the sample means around their stratum means, using the fixed external proportions. When $q_j=p_j^*$, the conventional variance includes an additional between-stratum component, giving a conservative test as shown in Remark~\ref{rem:conventional-conservative}.

The following result applies to the score and proposed new variance estimator in Equation~\eqref{eq:cae-survival-statistic}.

\begin{theorem}
\label{thm:tte}
Under the sampling model above, Assumption~\ref{ass:finite-observation} and $H_0$,
\[
  \frac{U}{\sqrt m}\xrightarrow{d}\mathcal N(0,V),
  \qquad
  \frac{\widehat V_{\mathrm{new}}}{m}\xrightarrow{p}V,
  \qquad
  V=\sum_j\frac{(p_j^*)^2v_j}{q_j}>0.
\]
Consequently, $T_{\mathrm{new}}\xrightarrow{d}\mathcal N(0,1)$.
\end{theorem}

We allow the limiting enrolled proportions $q_j$ to differ from the targets $p_j^*$. The calibrated inference result also applies to ordinary i.i.d. enrollment followed by MAIC matching the target stratum proportions, with $q_j$ equal to the candidate stratum probabilities. It therefore provides a common analysis for ordinary and adaptive enrollment.

For the conventional variance estimator, write $e_j=\E\{\Lambda_0(Y_k)\mid s_k=j\}$. The following proposition characterizes the difference between the asymptotic variance and the conventional variance estimator.

\begin{proposition}
\label{prop:conventional-limit}
Under the sampling model above and Assumption~\ref{ass:finite-observation}, with stratum weights $w_j=mp_j^*/n_j$, the estimator in Equation~\eqref{eq:survival-naive-var} satisfies
\begin{align}
  \frac{\widehat V_{\mathrm{conv}}}{m}
  &\ \xrightarrow{p}\ V_{\mathrm{conv}}
  =\sum_j\frac{(p_j^*)^2e_j}{q_j},
  \label{eq:conventional-limit}\\
  V_{\mathrm{conv}}-V
  &=\sum_j\frac{(p_j^*)^2}{q_j}(e_j-v_j).
  \label{eq:conventional-gap}
\end{align}
If $H_0$ also holds, then
\begin{equation}
  \sum_jp_j^*(e_j-v_j)=\sum_jp_j^*\mu_j^2.
  \label{eq:mixture-second-moment}
\end{equation}
\end{proposition}

\begin{remark}
\label{rem:conventional-conservative}
Under the conditions of Theorem~\ref{thm:tte}, suppose $q_j=p_j^*$ for every stratum. Then
\[
  V_{\mathrm{conv}}-V=\sum_jp_j^*\mu_j^2\ge0
  \quad\text{and}\quad
  \frac{U}{\sqrt{\widehat V_{\mathrm{conv}}}}
  \ \xrightarrow{d}\ \mathcal N(0,V/V_{\mathrm{conv}}).
\]
As a consequence, for any nominal level $0<\alpha<1$, the two-sided test that rejects when $|U|/\sqrt{\widehat V_{\mathrm{conv}}}>z_{1-\alpha/2}$ has limiting rejection probability less than $\alpha$ unless $\mu_j=0$ in every stratum. For other limiting proportions, the conventional one-sample log-rank test may not be conservative. Related conservativeness results arise under covariate-adaptive randomization, where covariates are balanced between treatment arms \citep{bugni2018inference,ye2020robust}.
\end{remark}

Section~\ref{sec:numerical} examines additional survival settings with more test statistics. Appendix~\ref{sec:supp-continuous} further gives the theoretical and empirical results under continuous endpoints where conventional variance estimators can also be conservative.

\section{Additional Survival Simulations\label{sec:numerical}}

In this section, we use additional simulations to examine how covariate shifts between the SAT and EC affect effective sample size and test performance. We vary the distributions of prognostic and predictive factors, comparing CAE with ordinary enrollment followed by MAIC or IPW.

We simulated SAT cohorts with fixed enrolled size $n_{\mathrm{SAT}}=150$ and a fixed EC with 200 patients. A fixed pooled survival benchmark is calibrated to the empirical 1.5-year survival of 0.49. Factor $A$ is prognostic, and $B$ modifies the treatment effect under $H_1$. The four candidate populations have $\{\Pr(A=1),\Pr(B=1)\}$ equal to $(0.70,0.30)$, $(0.50,0.30)$, $(0.70,0.50)$ and $(0.50,0.50)$, denoted No Shift, Prognostic Shift, Predictive Shift and Both Shift. The fixed EC's observed margins are 0.660 and 0.305. Each setting uses 5,000 replicates per hypothesis.

CAE and MAIC use the EC marginal covariate summaries and pooled survival benchmark. Inverse probability weighting (IPW) additionally uses the external joint cell proportions, computed from the fixed EC in this evaluation. It therefore represents access to more covariate information when both factors are observed. CAE-conventional and CAE-new assess variance choice for OSLRT within the same CAE cohort. We compare CAE, MAIC and IPW using jackknife analyses of OSLRT, the modified one-sample log-rank test (MOSLRT), the landmark Kaplan--Meier test (LMKM) and the HR test. MOSLRT denotes the one-sample log-rank test proposed in \cite{wu2014new}, and HR uses the log observed-to-expected event ratio, $\log(O/E)$. Appendix~\ref{sec:supp-survival-sim} gives the formulas and design details. All tests use two-sided normal critical values at nominal level 0.05.

\begin{figure}[!htbp]
\centering
\includegraphics[width=\linewidth]{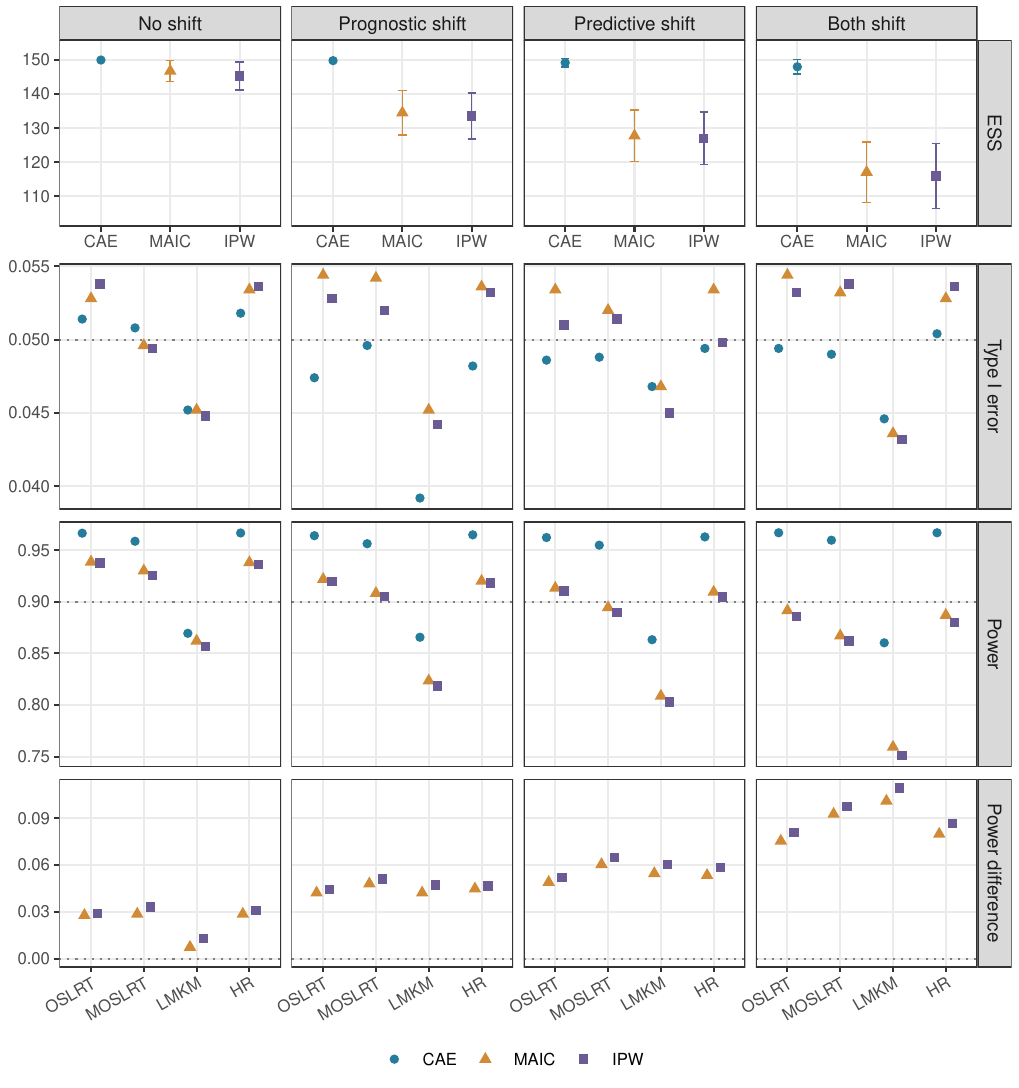}
\caption{Additional survival simulations. Each column corresponds to one candidate population. Rows show ESS, type I error, power, and power difference, from top to bottom. ESS is summarized by its mean with error bars of $\pm1$ SD. Power differences are calculated as CAE minus MAIC or IPW where positive values indicate higher power for CAE. All tests use jackknife inference.}
\label{fig:survival-ess}
\label{fig:survival-operating}
\label{fig:survival-jk-difference}
\end{figure}

Figure~\ref{fig:survival-operating} summarizes ESS, rejection probabilities under $H_0$ and $H_1$, and the power differences between strategies under $H_1$. Across the four main scenarios, CAE-JK OSLRT has $H_0$ rejection probabilities between 0.047 and 0.051. CAE-conventional and CAE-new give similar rejection probabilities in these scenarios and are omitted from the plot.

CAE reduces the covariate shift between SAT and EC, and its ESS remains close to the enrolled sample size across the four scenarios. Ordinary MAIC and IPW have lower ESS as the covariate shift level increases, particularly under Both Shift. These results show that adjusting population composition during enrollment can reduce the loss of ESS after weighting. Under Both Shift, CAE-JK and MAIC-JK OSLRT have $H_1$ rejection probabilities of 0.967 and 0.891, with $H_0$ probabilities of 0.049 and 0.054. IPW-JK gives 0.886 under $H_1$ and 0.053 under $H_0$. Appendix~\ref{sec:supp-unmeasured} examines sensitivity to omitting an important prognostic or predictive factor.

\section{Discussion\label{sec:discussion}}

In this paper, we have developed covariate-adaptive enrollment to address covariate shift between a single-arm trial population and its external reference during recruitment. We establish conditions for strong balance in the selected covariate summaries, regardless of endpoint type. For the weighted one-sample log-rank test, we provide a consistent variance estimator and characterize when the conventional variance estimator is conservative.

The DFMO evaluation illustrates the practical trade-off: compared with ordinary enrollment followed by retrospective weighting analysis with jackknife inference, prospective adaptive enrollment gives a higher effective sample size and higher statistical power, with additional screening and a longer trial duration. One extension would be to enroll and treat all eligible candidates, using the full cohort to estimate absolute outcomes, such as survival probabilities, and a prospectively selected CAE subcohort for comparison with the external control. Patients not selected by CAE could then remain in the broader study. Another direction is to adapt the biased-coin probabilities as enrollment goes on, to balance population alignment against enrollment speed.

Choosing an appropriate external reference remains important (see the review by \citet{zhu2026ectreview}). The simulations also show that unmeasured confounding can inflate type I error even after balancing observed covariates (Appendix~\ref{sec:supp-unmeasured}). Further work could account for uncertainty in the external reference and extend the methods to multiple external data sources.

\section*{Acknowledgements}
JX thanks the MRC Biostatistics Unit, University of Cambridge, and the Isaac Newton Institute for Mathematical Sciences for their hospitality during his participation in the programme \emph{Causal inference: From theory to practice and back again}. This work builds on a research project conducted during JX's internship at Astellas Pharma Inc. Earlier versions were presented at internal Astellas seminars, as a talk at \emph{Causal inference in the real world}, and as a poster at \emph{Causal inference in biomedicine}. JX thanks the participants at these events for helpful discussions. A recording of the talk is available at \url{https://gateway.newton.ac.uk/seminar/50329/}. JX also wishes to honor the memory of Professor Robin Evans, who chaired the session in which the talk was given.

\bibliography{references}

\clearpage
\appendix
\section*{Appendices}

\section{Supporting Results with a Survival Endpoint}
\label{sec:supp-survival-sim}

\subsection{DFMO Case Study}

The pseudo-IPD comprise 139 approximate records: 100 in Stratum 1 and 39 in Stratum 2. A pooled flexible parametric spline with $k=1$ internal knot supplies $S_0(t)=\exp\{-\Lambda_0(t)\}$. Its cumulative hazard is rescaled to give
\[
 S_0(2)=\frac{100\times0.84+39\times0.51}{139}.
\]
The fitted benchmark is held fixed across replicates.

To construct a stratum model compatible with this curve, set $p=100/139$ and $r=\log(0.51)/\log(0.84)$. Define $A(t)$ as the nonnegative solution of
\[
 p\exp\{-A(t)\}+(1-p)\exp\{-rA(t)\}=S_0(t).
\]
Under $H_0$, the stratum survival functions are $\exp\{-A(t)\}$ and $\exp\{-rA(t)\}$. Under $H_1$, they are $\exp\{-0.60A(t)\}$ and $\exp\{-0.40rA(t)\}$. Thus survival under $H_0$, averaged using the external stratum proportions, matches the analysis benchmark, and the two $H_1$ survival probabilities at two years are approximately 0.901 and 0.764.

Candidates have independent stratum labels with probability 0.5 for each stratum. Starting with zero imbalance, CAE accepts all candidates until both strata have been observed. Subsequent acceptance probabilities are 1.00, 0.50 and 0.75 when enrollment decreases, increases or leaves unchanged the squared cumulative imbalance defined in Section~\ref{sec:methodology}, respectively. Candidate interarrival times are independent exponential variables with rate 50 per year. CAE stops when 100 patients are enrolled. The ordinary cohort comprises the first 100 candidates from the same arrival sequence. Each cohort is analyzed two years after its own last patient enters.

In both weighted cohorts, the weights are $w_i=100p_j^*/n_j$ for subjects in stratum $j$. Each hypothesis uses 5,000 independent replicates. Within each replicate, $H_0$ and $H_1$ share the same enrollment design.

\subsection{Additional Survival Simulations}

The fixed EC contains 200 patients, with observed proportions 0.660 for $A=1$ and 0.305 for $B=1$. Its pooled survival benchmark is $S_0(t)=\exp\{-\Lambda_0(t)\}$, calibrated so that $S_0(1.5)=0.49$.

We generate SAT event times from a proportional hazards model, with $A$ as a prognostic factor and $B$ as a predictive factor. Let $L(t)$ be the baseline cumulative hazard for $A=0$. Under $H_0$, the conditional cumulative hazard satisfies
\begin{equation}
 \log\Lambda_{\mathrm{SAT}}(t\mid A,B)
 =
 \log L(t)+\beta_A A,
 \qquad t>0.
 \label{eq:survival-dgm-control}
\end{equation}
Under $H_1$, we add a treatment effect that depends on $B$:
\begin{equation}
 \log\Lambda_{\mathrm{SAT}}(t\mid A,B)
 =
 \log L(t)+\beta_A A+\log(0.70)-0.10B,
 \qquad t>0.
 \label{eq:survival-dgm-sat}
\end{equation}
The prognostic coefficient is $\beta_A=0.489$ ($r_A=1.631$), estimated by fitting a Cox model with covariate $A$ to the fixed EC. This coefficient is held fixed under both hypotheses. Under $H_1$, the hazard is multiplied by $0.70\exp(-0.10B)$ relative to $H_0$.

We choose $L(t)$ so that the survival distribution under $H_0$, averaged using the external proportion of $A$, equals the pooled benchmark:
\[
 (1-0.660)\exp\{-L(t)\}
 +0.660\exp\{-L(t)\exp(\beta_A)\}
 =
 S_0(t).
\]
Thus the generating model includes the prognostic differences between patients while preserving the pooled reference under $H_0$. The survival tests use the pooled cumulative hazard $\Lambda_0(t)$.

The enrolled sample size is fixed at 150. Candidate factors $A$ and $B$ are independent Bernoulli variables with probabilities given in Table~\ref{tab:survival-scenarios}.

\begin{table}[htbp]
\centering
\caption{Candidate distributions and measured adjustment sets in the additional survival simulations. The No Shift probabilities are 0.70 and 0.30; the corresponding observed margins in the fixed EC are 0.660 and 0.305.}
\label{tab:survival-scenarios}
\begin{tabular}{lccc}
\toprule Population & $\Pr(A=1)$ & $\Pr(B=1)$ & Measured factors \\ \midrule No shift         & 0.70 & 0.30 & $A,B$ \\ Prognostic shift & 0.50 & 0.30 & $A,B$ \\ Predictive shift & 0.70 & 0.50 & $A,B$ \\ Both shift       & 0.50 & 0.50 & $A,B$ \\ Omit B           & 0.50 & 0.50 & $A$ only \\ Omit A           & 0.50 & 0.50 & $B$ only \\ \bottomrule
\end{tabular}
\end{table}

When both factors are observed, CAE initially accepts every candidate until all four joint $A,B$ cells have appeared. In the omission scenarios with unmeasured confounding, initialization ends when both levels of the observed factor have appeared. Only the observed factors are used in subsequent enrollment and analysis. After initialization, acceptance probabilities are 0.90, 0.30 and 0.60 for improvement, worsening and a tie, respectively. Initialization patients contribute to the cumulative imbalance and final analysis.

CAE stops when 150 patients are enrolled. Both strategies use the same candidate arrival sequence at rate 150 per year. The ordinary cohort keeps the first 150 candidates. Each cohort is analyzed 1.5 years after its own final entry. Each setting uses 5,000 independent replicates per hypothesis. MAIC calibrates the measured marginal proportions. With both factors observed, it need not reproduce the external joint cell proportions. IPW instead weights each observed cell to its external proportion.

Weights are normalized so that $\sum_iw_i=m$, where $m$ is the enrolled sample size. Write
\[
 O=\sum_iw_i\Delta_i,
 \qquad
 E=\sum_iw_i\Lambda_0(Y_i),
 \qquad
 O_2=\sum_iw_i^2\Delta_i,
 \qquad
 E_2=\sum_iw_i^2\Lambda_0(Y_i).
\]
The conventional OSLRT statistic is $(O-E)/\sqrt{E_2}$, and the MOSLRT statistic is $(O-E)/\sqrt{(O_2+E_2)/2}$. The conventional HR statistic divides $\log(O/E)$ by $\sqrt{O_2}/O$. CAE-new uses the variance estimator $\sum_j n_jw_j^2s_j^2$, with strata defined by the joint values of the measured factors. Here $n_j$ is the stratum size, $w_j$ is the common within-stratum weight from marginal MAIC calibration, and $s_j^2$ is the sample variance of the score contributions $M_i$, with divisor $n_j-1$.

At landmark $t_L$, LMKM compares weighted SAT survival with $S_0(t_L)$. Here $t_L=2$ years for DFMO and 1.5 years for the additional survival study. No patient is censored before the landmark. The weighted Kaplan--Meier estimate therefore equals
\[
 \hat S=m^{-1}\sum_iw_i I_i,
 \qquad
 I_i=1\{F_i>t_L\}.
\]
Its conventional variance estimator is
\[
 \widehat V_{\mathrm{LMKM}}
 =m^{-2}\sum_iw_i^2(I_i-\hat S)^2,
\]
and the conventional statistic is $(\hat S-S_0(t_L))/\sqrt{\widehat V_{\mathrm{LMKM}}}$. For each test, jackknife inference follows Section~\ref{sec:inference}, re-estimating the weights and recomputing the conventional statistic after each deletion.

\subsection{Survival Sensitivity to Unmeasured Confounding}
\label{sec:supp-unmeasured}

We examine sensitivity to unmeasured confounding in two scenarios, both using the candidate distribution of the Both Shift scenario. In the first, the prognostic factor $A$ is observed but the predictive factor $B$ is unmeasured. In the second, $B$ is observed but $A$ is unmeasured. Enrollment and adjustment use only the observed factor, while outcomes are generated from Equation~\eqref{eq:survival-dgm-control} under $H_0$ and Equation~\eqref{eq:survival-dgm-sat} under $H_1$.

\begin{figure}[!htbp]
\centering
\includegraphics[width=\linewidth]{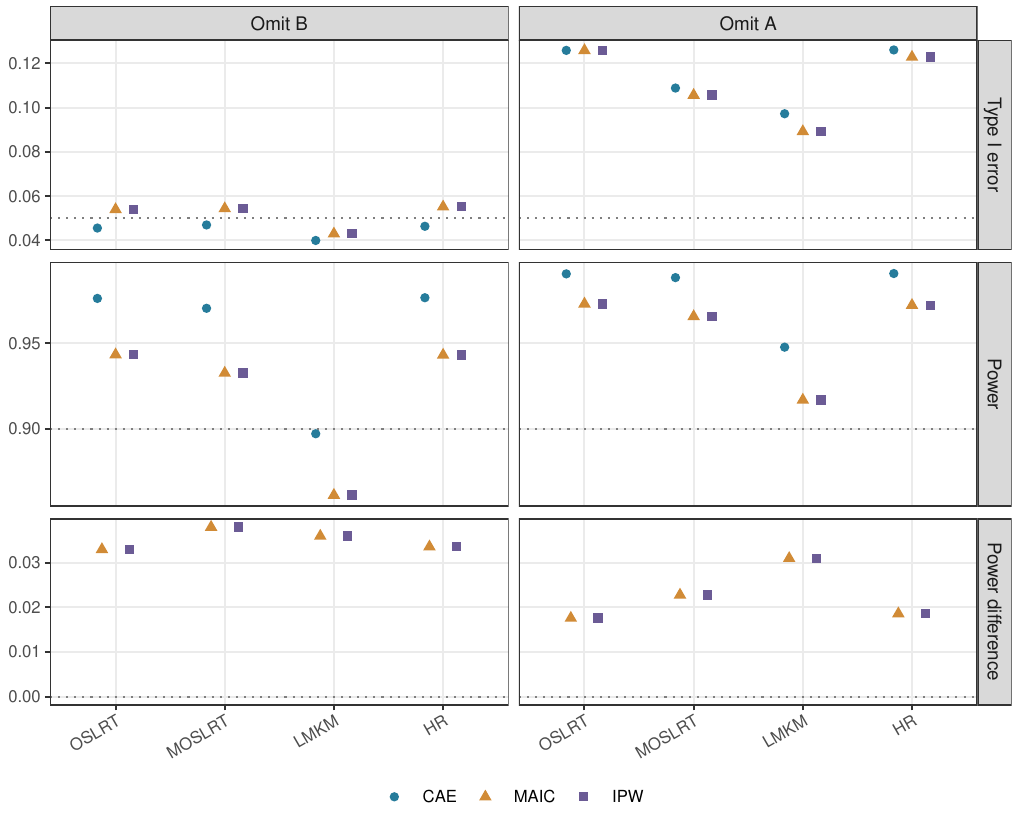}
\caption{Survival simulations with an omitted covariate. Each column corresponds to one omission scenario: Omit B excludes the predictive factor from enrollment and adjustment, whereas Omit A excludes the prognostic factor. Rows show type I error, power, and power difference, from top to bottom. Power differences are calculated as CAE minus MAIC or IPW, with positive values indicating higher power for CAE. All tests use jackknife inference.}
\label{fig:surv-unmeasured-operating}
\label{fig:omitted-jk-difference}
\end{figure}

Figure~\ref{fig:surv-unmeasured-operating} displays the jackknife analyses for both unmeasured confounding settings. For CAE-JK OSLRT, $H_0$ rejection is 0.126 when the prognostic factor $A$ is omitted, compared with 0.046 when only the predictive factor $B$ is omitted. Enrollment and calibration use only the observed factor and cannot recover prognostic information that is absent from the adjustment set. $H_1$ rejection in these settings is interpreted alongside the $H_0$ results.

\section{Supporting Results with a Continuous Endpoint}
\label{sec:supp-continuous}

We extend the analysis to a continuous endpoint and evaluate CAE using a simulation model based on the CBASP trial \citep{keller2000comparison}.

\subsection{Continuous Endpoint Inference}

For continuous outcomes, we use regression calibration to adjust the enrolled outcome mean to the EC covariate means. As before, the EC summaries are treated as fixed. Let $\boldsymbol B=\boldsymbol b(X)$ be the features balanced by CAE, with target mean $\boldsymbol\mu_{EC}=\E_{EC}(\boldsymbol B)$. Here the design covariate vector is $\boldsymbol x=\boldsymbol B$. Throughout this subsection, $\E_{SAT}$ denotes expectation under the candidate distribution, and $Y_i=Y_i(1)$ is the observed continuous outcome for an enrolled patient. The following assumption describes the outcome model and the enrollment mechanism used for this analysis.

\begin{assumption}
\label{ass:cont-reg}
Candidate records $(\boldsymbol B_i,Y_i(1))$ are i.i.d. Let $U_i$ be independent $\mathrm{Unif}(0,1)$ variables, independent of all candidate records. The adaptive enrollment rule uses only previous covariates and enrollment decisions, the current $\boldsymbol B_i$, and $U_i$, with $Z_i=1\{U_i\le\phi(\boldsymbol D_{i-1},\boldsymbol B_i)\}$. To describe the relationship between the outcome and the balanced covariates, define the linear projection
\[
 (\alpha_*,\boldsymbol\beta_*)
 =
 \arg\min_{\alpha,\boldsymbol\beta}
 \E_{SAT}\{Y(1)-\alpha
 -\boldsymbol\beta^\top(\boldsymbol B-\boldsymbol\mu_{EC})\}^2,
\]
with residual
\[
 \varepsilon_*
 =
 Y(1)-\alpha_*
 -\boldsymbol\beta_*^\top(\boldsymbol B-\boldsymbol\mu_{EC}).
\]
Assume $\E_{SAT}(\varepsilon_*\mid\boldsymbol B)=0$ and $\E_{SAT}|\varepsilon_*|^{2+\eta}<\infty$ for some $\eta>0$. Let $P_B$ be the candidate distribution of $\boldsymbol B$, and write $h(\boldsymbol b)= \E_{SAT}(\varepsilon_*^2\mid\boldsymbol B=\boldsymbol b)$ for the conditional residual variance.

Because enrollment depends on the covariates, the relevant moments are those of the enrolled population. Let $\pi$ be the invariant distribution of the imbalance chain and $\phi$ the enrollment probability in \eqref{eq:design_function}. Define the limiting enrollment fraction
\[
 \rho_\pi
 =
 \int\int \phi(\boldsymbol d,\boldsymbol b)
 P_B(d\boldsymbol b)\pi(d\boldsymbol d)
\]
and the corresponding average residual variance
\[
 \sigma_{\mathrm{cal}}^2
 =
 \rho_\pi^{-1}
 \int\int \phi(\boldsymbol d,\boldsymbol b)h(\boldsymbol b)
 P_B(d\boldsymbol b)\pi(d\boldsymbol d)
 \in(0,\infty).
\]
The feature dimension is fixed, and the limiting regression moment matrix
\[
 Q
 =
 \rho_\pi^{-1}
 \int\int \phi(\boldsymbol d,\boldsymbol b)
 (1,\boldsymbol b^\top)^\top(1,\boldsymbol b^\top)
 P_B(d\boldsymbol b)\pi(d\boldsymbol d)
\]
is positive definite.
\end{assumption}

The residual condition specifies a linear conditional mean while allowing the residual variance to depend on the covariates. For $N$ screened candidates, let $\mathcal I_N=\{i\le N:Z_i=1\}$ be the enrolled set and $R_N=|\mathcal I_N|$ its size. We first establish the limits of the enrolled sample moments and the residual mean, which are needed for regression calibration.

\begin{lemma}
\label{lemma:stationary-enrolled-limits}
Under the conditions of Theorem~\ref{thm:balance} and Assumption~\ref{ass:cont-reg},
\[
 \frac{R_N}{N}\xrightarrow{p}\rho_\pi,
 \qquad
 R_N^{-1}\sum_{i\in\mathcal I_N}
 (1,\boldsymbol B_i^\top)^\top(1,\boldsymbol B_i^\top)
 \xrightarrow{p}Q,
\]
and
\[
 R_N^{-1}\sum_{i\in\mathcal I_N}\varepsilon_{*,i}^2
 \xrightarrow{p}\sigma_{\mathrm{cal}}^2,
 \qquad
 R_N^{-1/2}\sum_{i\in\mathcal I_N}\varepsilon_{*,i}
 \xrightarrow{d}\mathcal N(0,\sigma_{\mathrm{cal}}^2).
\]
\end{lemma}

\begin{proof}
We first consider averages of functions of the enrolled covariates. Let
\[
 \mathcal F_i
 =\sigma\{\boldsymbol B_k,U_k,Y_k(1):k\le i\},
 \qquad
 \mathcal G_i
 =\mathcal F_{i-1}\vee\sigma(\boldsymbol B_i,U_i).
\]
For an integrable function $g$, define
\[
 a_g(\boldsymbol d)
 =
 \int\phi(\boldsymbol d,\boldsymbol b)g(\boldsymbol b)
 P_B(d\boldsymbol b).
\]
Independence of the current candidate record and $U_i$ from the past gives
\[
 \E\{Z_i g(\boldsymbol B_i)\mid\mathcal F_{i-1}\}
 =a_g(\boldsymbol D_{i-1}).
\]
Thus each contribution can be separated into its conditional mean and a martingale difference. The ergodic theorem for the positive Harris recurrent chain, started at $\boldsymbol D_0=0$, yields
\[
 N^{-1}\sum_{i=1}^N a_g(\boldsymbol D_{i-1})
 \xrightarrow{p}\int a_g\,d\pi.
\]
For bounded $g$, the average of $Z_i g(\boldsymbol B_i)-a_g(\boldsymbol D_{i-1})$ converges to zero by the martingale variance bound. For integrable $g$, truncation at $K$ leaves expected tail contributions bounded by $2\E\{|g(\boldsymbol B)|1(|g(\boldsymbol B)|>K)\}$, uniformly in $N$. Consequently,
\[
 N^{-1}\sum_{i=1}^N Z_i g(\boldsymbol B_i)
 \xrightarrow{p}\int a_g\,d\pi.
\]
Taking $g=1$ gives the enrollment-fraction limit. Applying the same result to the entries of $(1,\boldsymbol B^\top)^\top(1,\boldsymbol B^\top)$ and dividing by $R_N/N$ gives the limit $Q$.

We next apply this argument to the residual variance, using $g=h$. The difference between the squared residual and its conditional mean satisfies
\[
 \E\{Z_i(\varepsilon_{*,i}^2-h(\boldsymbol B_i))
       \mid\mathcal G_i\}=0.
\]
The $(2+\eta)$ moment condition and a martingale moment inequality imply that the average of these differences converges to zero in probability. Dividing by $R_N/N\xrightarrow{p}\rho_\pi>0$ therefore gives the residual second-moment limit.

It remains to establish the normal limit for the residual sum. Let $\xi_i=Z_i\varepsilon_{*,i}$. Then
\[
 \E(\xi_i\mid\mathcal G_i)=0,
 \qquad
 \E(\xi_i^2\mid\mathcal G_i)=Z_i h(\boldsymbol B_i),
 \qquad
 \E(\xi_i^2\mid\mathcal F_{i-1})
 =a_h(\boldsymbol D_{i-1}).
\]
Hence $(\xi_i)$ is a martingale difference sequence whose predictable variance satisfies
\[
 N^{-1}\sum_{i=1}^N
 \E(\xi_i^2\mid\mathcal F_{i-1})
 \xrightarrow{p}\rho_\pi\sigma_{\mathrm{cal}}^2.
\]
The residual moment condition also gives the conditional Lindeberg condition: for every $\epsilon>0$,
\[
 \frac1N\sum_{i=1}^N
 \E\{\xi_i^2 1(|\xi_i|>\epsilon\sqrt N)
       \mid\mathcal F_{i-1}\}
 \le
 \frac{\E|\varepsilon_*|^{2+\eta}}
      {\epsilon^\eta N^{\eta/2}}
 \longrightarrow0.
\]
The martingale central limit theorem and $R_N/N\xrightarrow{p}\rho_\pi$ give the final limit.
\end{proof}

We now use the fitted regression to adjust the enrolled outcome mean to the EC covariate means. Under the linear conditional mean model, this adjustment targets $\alpha_*$, which is compared with the fixed EC outcome mean $\bar Y_{EC}(0)$. The next theorem gives the resulting null distribution and variance estimator.

\begin{theorem}
\label{thm:cont}
Under the conditions of Theorem~\ref{thm:balance}, Assumption~\ref{ass:cont-reg}, and $H_0:\alpha_*=\bar Y_{EC}(0)$, define
\[
 \hat\theta_{\mathrm{cal}}
 =
 \bar Y_{SAT}
 -
 \hat{\boldsymbol\beta}^{\top}
 (\bar{\boldsymbol B}_{SAT}-\boldsymbol\mu_{EC}),
\]
where $\hat{\boldsymbol\beta}$ is the ordinary least squares slope from the enrolled regression of $Y$ on $1$ and $\boldsymbol B$, and $\bar{\boldsymbol B}_{SAT} =R_N^{-1}\sum_{i\in\mathcal I_N}\boldsymbol B_i$. Then
\[
 \sqrt{R_N}
 \{\hat\theta_{\mathrm{cal}}-\bar Y_{EC}(0)\}
 \xrightarrow{d}\mathcal N(0,\sigma_{\mathrm{cal}}^2).
\]
With fitted residuals $\hat\varepsilon_i$ and $q=\dim(\boldsymbol B)$, let
\[
 \hat\sigma_{\mathrm{cal}}^2
 =
 (R_N-q-1)^{-1}
 \sum_{i\in\mathcal I_N}\hat\varepsilon_i^2.
\]
Then
\[
 \frac{\sqrt{R_N}
 \{\hat\theta_{\mathrm{cal}}-\bar Y_{EC}(0)\}}
 {\hat\sigma_{\mathrm{cal}}}
 \xrightarrow{d}\mathcal N(0,1).
\]
\end{theorem}

Strong balance gives $\bar{\boldsymbol B}_{SAT}-\boldsymbol\mu_{EC}=O_p(N^{-1})$. The remaining covariate imbalance is smaller than the fluctuation of the residual mean. The first-order uncertainty is therefore determined by the residual variation among enrolled patients.

\subsection{Continuous Endpoint Simulation}

We examine how covariate shift affects effective sample size and test performance for a continuous endpoint. As in the survival study, we vary the candidate population while keeping the EC fixed and compare CAE with ordinary enrollment followed by weighting.

We simulated SAT cohorts of 135 enrolled patients using a fixed synthetic EC with 222 patients. The EC was constructed from the combination-treatment records of a trial comparing nefazodone, the cognitive behavioral-analysis system of psychotherapy (CBASP), and their combination for chronic depression \citep{keller2000comparison}. The outcome is the post-treatment 24-item Hamilton Rating Scale for Depression score, denoted $\text{FinalHAMD}$. The baseline covariates are age and the baseline score, denoted $\text{HAMD24}$.

We use these records to fit the EC outcome model
\begin{equation}
\label{eq:additive_model}
 \text{FinalHAMD}_i(0)
 =
 \beta_0+\beta_1\text{Age}_i
 +\beta_2\text{HAMD24}_i+\tilde\varepsilon_i.
\end{equation}
Independent Student-$t$ errors with two degrees of freedom are added to the fitted values, and the resulting outcomes are rounded to construct the fixed EC. Refitting the regression supplies the SAT generating coefficients. The EC mean age is 44.30 years, and its mean baseline HAMD24 score is 27.38.

To vary the candidate composition, we generate age and baseline HAMD24 independently from rounded normal distributions with SDs 6 and 4. No Shift uses the EC means; Age Shift changes mean age to 41 years; HAMD24 Shift changes the mean baseline score to 29; Both Shift changes both. CAE accepts the first ten candidates and includes them in the cumulative imbalance. Subsequent acceptance probabilities are 0.90, 0.30 and 0.60 when enrollment improves, worsens or leaves the imbalance unchanged, respectively.

Each setting uses 5,000 replicates per hypothesis. Under $H_0$, SAT outcomes follow
\begin{equation}
 \text{FinalHAMD}_i(1)
 =
 \beta_0+\beta_1\text{Age}_i
 +\beta_2\text{HAMD24}_i+\varepsilon_i,
\end{equation}
where $\varepsilon_i\sim\mathcal N(0,25)$ independently. Under $H_1$, the coefficient of baseline HAMD24 is reduced by 0.05:
\begin{equation}
 \text{FinalHAMD}_i(1)
 =
 \beta_0+\beta_1\text{Age}_i
 +(\beta_2-0.05)\text{HAMD24}_i+\varepsilon_i.
\end{equation}

After enrollment, calibration weights align the measured covariate means with the EC targets. Let $m$ be the enrolled sample size and normalize the weights so that $\sum_iq_i=1$. Write $\bar Y_q=\sum_iq_iY_i$ and $\hat\theta_w=\bar Y_q-\bar Y_{EC}$. The conventional variance estimator is
\[
 \widehat V_{\mathrm{conv},w}
 =
 \frac{m}{m-1}\sum_iq_i^2(Y_i-\bar Y_q)^2.
\]
For CAE-cal, we instead use residuals from weighted least squares regression of $Y_i$ on an intercept and the two covariates centered at the EC means. With weights $q_i$ and fitted residuals $\hat e_i$, the variance estimator is
\[
 \widehat V_{\mathrm{cal},w}
 =
 \frac{m}{m-3}
 \left(\sum_iq_i\hat e_i^2\right)
 \left(\sum_iq_i^2\right)
 =
 \frac{m\sum_iq_i\hat e_i^2}
      {(m-3)\operatorname{ESS}}.
\]
Both analyses use $\hat\theta_w$ as the numerator. Theorem~\ref{thm:cont} concerns OLS regression calibration under the stated covariate model and a fixed number of screened candidates.

For ordinary enrollment, we consider MAIC and the covariate balancing propensity score (CBPS) method. MAIC uses the external covariate means. CBPS uses the exact average treatment effect on the treated specification \citep{imai2014covariate} and the fixed EC covariate records. Its weights are normalized jointly over EC and SAT observations, capped at 5, and then normalized within SAT. We compare CAE, MAIC and CBPS using jackknife inference. Each deletion re-estimates the weights and recomputes the conventional standardized contrast using the remaining SAT patients and the same EC records.

\begin{figure}[!htbp]
\centering
\includegraphics[width=\linewidth]{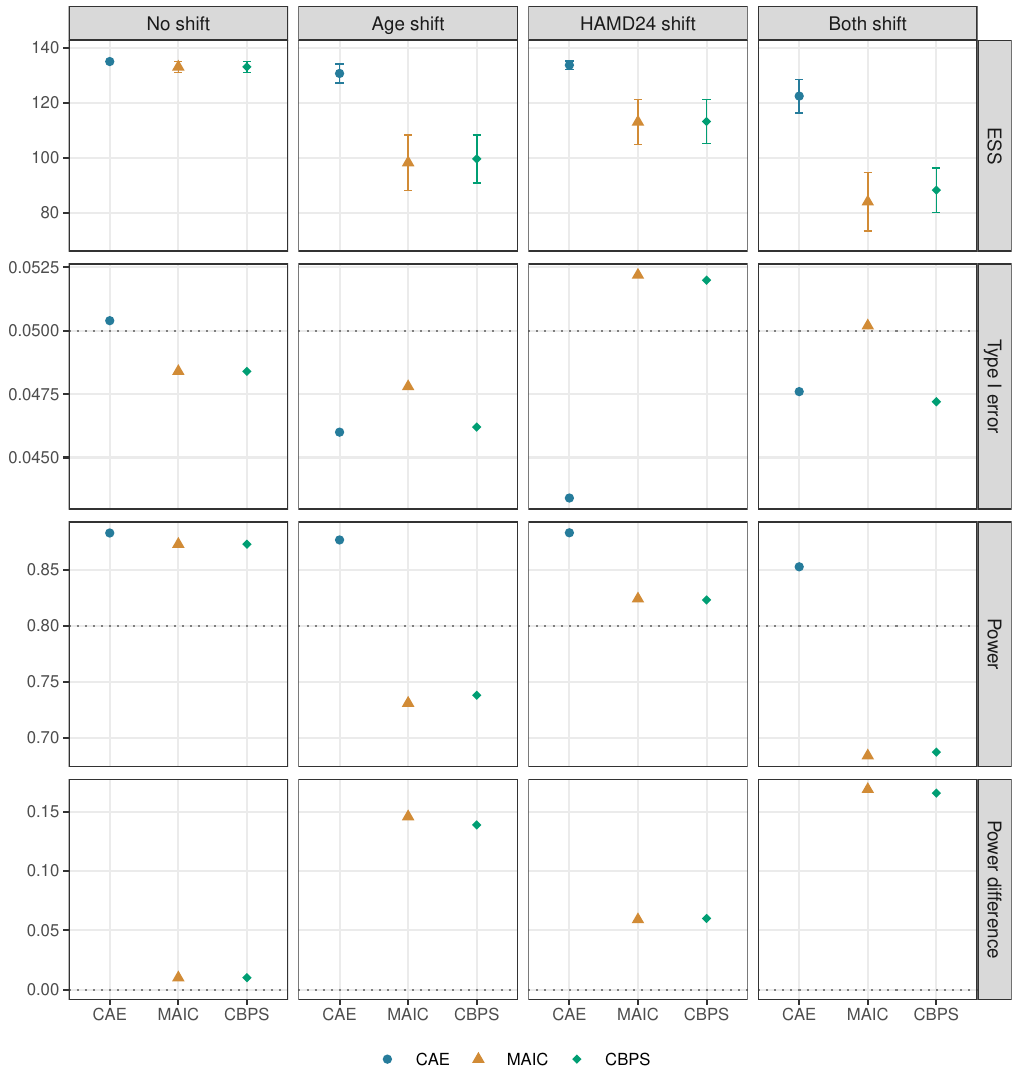}
\caption{Continuous endpoint simulations. Each column corresponds to one candidate population. Rows show ESS, type I error, power, and power difference, from top to bottom. ESS is summarized by its mean with error bars of $\pm1$ SD. Power differences are calculated as CAE minus MAIC or CBPS, with positive values indicating higher power for CAE. All tests use jackknife inference.}
\label{fig:continuous-ess}
\label{fig:continuous-operating}
\label{fig:continuous-jk-difference}
\end{figure}

Figure~\ref{fig:continuous-operating} summarizes ESS, type I error, power and power differences between enrollment strategies. CAE keeps ESS close to the enrolled sample size across the four scenarios. Ordinary enrollment followed by weighting has lower ESS as the candidate means depart from the EC targets. As in the survival study, adjusting composition during enrollment reduces the loss of ESS after weighting.

Within the CAE cohort, CAE-cal has $H_0$ rejection probabilities between 0.048 and 0.054, compared with 0.037 to 0.043 for CAE-conventional. With jackknife inference, the corresponding ranges are 0.043 to 0.050 for CAE and 0.046 to 0.052 for CBPS.

\section{Supporting Proofs}

\subsection{Markov-chain stability for the imbalance process}
\label{app:ergodicity}

\begin{proposition}[Ergodicity]
\label{prop:ergodicity}
Under the screened-candidate setup above, suppose Assumption~\ref{ass:covariates} holds, and either Assumption~\ref{ass:continuous} or Assumption~\ref{ass:discrete} holds. Let $\boldsymbol{u}=\boldsymbol{x}-\boldsymbol{\mu}_{EC}$, $\boldsymbol{\delta}=\E[\boldsymbol{u}]$, and
\[
c_*=\inf_{\|\boldsymbol{v}\|=1}\E\{|\boldsymbol{v}^\top \boldsymbol{u}|\}.
\]
Suppose $0<p_{\mathrm{avg}}<1$ and the biasing parameter $\Delta p$ is feasible for the biased-coin probability, that is,
\[
    0<\Delta p<2\min\{p_{\mathrm{avg}},1-p_{\mathrm{avg}}\}.
\]
If $\Delta p$ also satisfies the drift condition
\begin{equation}
    \Delta p > \frac{2p_{\mathrm{avg}} \|\boldsymbol{\delta}\|}{c_*},
\end{equation}
then the imbalance process $\{\boldsymbol{D}_n\}$ satisfies the Foster--Lyapunov drift condition and the required small-set condition. Specifically, there exists a Lyapunov function $V(\boldsymbol{D}) = \|\boldsymbol{D}\|^2$ and constants $\epsilon > 0, K < \infty$ such that $\E[V(\boldsymbol{D}_{n+1}) - V(\boldsymbol{D}_n) \mid \mathcal{F}_n] \le -\epsilon$ whenever $\|\boldsymbol{D}_n\| > K$, and any bounded set in the state space is a small set in the continuous case or a petite set in the discrete case. Consequently, by standard drift-minorization theory, the Markov chain $\{\boldsymbol{D}_n\}$ is positive Harris recurrent and ergodic on its reachable state space, possesses a unique stationary distribution $\pi$, and satisfies $\int \|\boldsymbol{d}\|\,\pi(d\boldsymbol{d})<\infty$.
\end{proposition}

The proof of Proposition~\ref{prop:ergodicity} requires verifying two conditions: the Foster--Lyapunov drift condition and the small-set (or petite-set) condition. We establish both under the continuous and discrete covariate assumptions.

\subsubsection{Verification of Drift Condition}
\label{app:drift}

Write $\boldsymbol u=\boldsymbol x-\boldsymbol\mu_{EC}$ and $\boldsymbol\delta=\E(\boldsymbol u)$. The enrollment probability is
\[
  \phi(\boldsymbol d,\boldsymbol x)
  =p_{\mathrm{avg}}
  -\frac{\Delta p}{2}
   \operatorname{sgn}
   \left(2\boldsymbol d^\top\boldsymbol u
         +\|\boldsymbol u\|^2\right).
\]
The strict feasibility condition in Proposition~\ref{prop:ergodicity} implies
\[
  \iota\le\phi(\boldsymbol d,\boldsymbol x)\le1-\iota,
  \qquad
  \iota=
  \min\left\{
  p_{\mathrm{avg}}-\frac{\Delta p}{2},
  1-p_{\mathrm{avg}}-\frac{\Delta p}{2}
  \right\}>0.
\]

Let $V(\boldsymbol d)=\|\boldsymbol d\|^2$ and write
\[
  \Delta V(\boldsymbol d)
  =(PV)(\boldsymbol d)-V(\boldsymbol d)
  =\E\left\{
  V(\boldsymbol D_{n+1})-V(\boldsymbol D_n)
  \mid\boldsymbol D_n=\boldsymbol d
  \right\}.
\]
Since $Z^2=Z$, conditioning on the candidate covariate gives
\begin{align*}
  \Delta V(\boldsymbol d)
  &=
  \E\left[
  \phi(\boldsymbol d,\boldsymbol x)
  \left(2\boldsymbol d^\top\boldsymbol u
        +\|\boldsymbol u\|^2\right)
  \right]\\
  &=
  2p_{\mathrm{avg}}\boldsymbol d^\top\boldsymbol\delta
  +p_{\mathrm{avg}}\E\|\boldsymbol u\|^2
  -\frac{\Delta p}{2}
   \E\left|
   2\boldsymbol d^\top\boldsymbol u+\|\boldsymbol u\|^2
   \right|.
\end{align*}
The reverse triangle inequality gives
\[
  \left|
  2\boldsymbol d^\top\boldsymbol u+\|\boldsymbol u\|^2
  \right|
  \ge
  2|\boldsymbol d^\top\boldsymbol u|-\|\boldsymbol u\|^2.
\]
Consequently,
\[
  \Delta V(\boldsymbol d)
  \le
  2p_{\mathrm{avg}}\|\boldsymbol d\|
                     \|\boldsymbol\delta\|
  -\Delta p\,\E|\boldsymbol d^\top\boldsymbol u|
  +\left(p_{\mathrm{avg}}+\frac{\Delta p}{2}\right)
   \E\|\boldsymbol u\|^2.
\]

Recall that
\[
  c_*=
  \inf_{\|\boldsymbol v\|=1}
  \E|\boldsymbol v^\top\boldsymbol u|.
\]
Because $\boldsymbol\mu_{EC}$ lies in the interior of the convex hull of the support of $\boldsymbol x$, $\boldsymbol u$ is not supported on any hyperplane through the origin. Thus $\E|\boldsymbol v^\top\boldsymbol u|>0$ for every unit vector $\boldsymbol v$. This expectation is continuous in $\boldsymbol v$, since
\[
  \left|
  \E|\boldsymbol v^\top\boldsymbol u|
  -\E|\boldsymbol w^\top\boldsymbol u|
  \right|
  \le
  \|\boldsymbol v-\boldsymbol w\|\,\E\|\boldsymbol u\|.
\]
Compactness of the unit sphere therefore gives $c_*>0$. It follows that
\[
  \E|\boldsymbol d^\top\boldsymbol u|
  \ge c_*\|\boldsymbol d\|.
\]

Under the stated drift condition, define
\[
  \kappa
  =\Delta p\,c_*
   -2p_{\mathrm{avg}}\|\boldsymbol\delta\|>0,
  \qquad
  b_0=
  \left(p_{\mathrm{avg}}+\frac{\Delta p}{2}\right)
  \E\|\boldsymbol u\|^2<\infty.
\]
Then, for every state $\boldsymbol d$,
\[
  \Delta V(\boldsymbol d)
  \le-\kappa\|\boldsymbol d\|+b_0.
\]
Choose $K>2b_0/\kappa$ and let $C$ be the set of reachable states with $\|\boldsymbol d\|\le K$. The preceding bound implies
\[
  (PV)(\boldsymbol d)-V(\boldsymbol d)
  \le
  -\frac{\kappa}{2}\|\boldsymbol d\|
  +b_0\mathbf 1_C(\boldsymbol d).
\]
In particular, $\Delta V(\boldsymbol d)\le-\epsilon$ outside $C$, where $\epsilon=\kappa K/2>0$. This proves the required drift bound.

\subsubsection{Small Set Condition\label{app:smallset}}

Based on the allocation mechanism, the imbalance process $\left\{ \boldsymbol{D}_n = \sum_{i=1}^{n} Z_i (\boldsymbol{x}_i - \boldsymbol{\mu}_{EC}) \right\}$ is a time-homogeneous Markov chain with state space $\mathbb{R}^k$ (under Assumption~\ref{ass:continuous}) or a countable subset of $\mathbb{R}^k$ (under Assumption~\ref{ass:discrete}). The transition probability kernel $P$ of the imbalance process $\left\{ \boldsymbol{D}_n \right\}$ can be defined through its action on bounded measurable functions $h$:
\begin{equation*}
    (P h)(\boldsymbol{d})
    = \int \left[
        \phi(\boldsymbol{d},\boldsymbol{t}) h(\boldsymbol{d}+\boldsymbol{t}-\boldsymbol{\mu}_{EC})
        + \left(1 - \phi(\boldsymbol{d},\boldsymbol{t})\right) h(\boldsymbol{d})
    \right] \Gamma(d\boldsymbol{t}) .
\end{equation*}
Equivalently, for any measurable set $A \subset \mathbb{R}^k$,
\begin{equation*}
    P(\boldsymbol{d},A)
    = P(\boldsymbol{d}, \mathbb{I}_A)
    = \int \left[
        \phi(\boldsymbol{d},\boldsymbol{t}) \mathbb{I}_A(\boldsymbol{d}+\boldsymbol{t}-\boldsymbol{\mu}_{EC})
        + \left(1 - \phi(\boldsymbol{d},\boldsymbol{t})\right) \mathbb{I}_A(\boldsymbol{d})
    \right] \Gamma(d\boldsymbol{t}) .
\end{equation*}
Under Assumption~\ref{ass:continuous}, the covariate has density $p(\boldsymbol{t})$, and the transition kernel reduces to
\begin{equation*}
    P(\boldsymbol{d},A)
    = \int_{\mathbb{R}^k}
    \left[
        \phi(\boldsymbol{d},\boldsymbol{t}) \mathbb{I}_A(\boldsymbol{d}+\boldsymbol{t}-\boldsymbol{\mu}_{EC})
        + \left(1 - \phi(\boldsymbol{d},\boldsymbol{t})\right) \mathbb{I}_A(\boldsymbol{d})
    \right]
    p(\boldsymbol{t}) d\boldsymbol{t} .
\end{equation*}
Under Assumption~\ref{ass:discrete}, the transition kernel becomes
\begin{equation*}
    P(\boldsymbol{d},A)
    = \sum_{i=1}^{s}
    \left[
        \phi(\boldsymbol{d},\boldsymbol{t}_i) \mathbb{I}_A(\boldsymbol{d}+\boldsymbol{t}_i-\boldsymbol{\mu}_{EC})
        + \left(1 - \phi(\boldsymbol{d},\boldsymbol{t}_i)\right) \mathbb{I}_A(\boldsymbol{d})
    \right]
    \mathbb{P}(\boldsymbol{x} = \boldsymbol{t}_i) .
\end{equation*}
Consequently, the imbalance process $\{\boldsymbol{D}_n\}$ is a time-homogeneous Markov chain with respect to the natural filtration $\mathcal{F}_n = \sigma(\boldsymbol{D}_1,\dots,\boldsymbol{D}_n)$, and its conditional transition probabilities satisfy
\begin{equation*}
    P( \boldsymbol{D}_{n+1} \in A \mid \mathcal{F}_n) = P( \boldsymbol{D}_{n+1} \in A \mid \boldsymbol{D}_n ) = P(\boldsymbol{D}_n, A) .
\end{equation*}

We present a lemma like the spread out condition in \cite{ma2024new}.

\begin{lemma}
\label{lemma:convolution-support}
Let $U$ have density $p_u\in L^2(\mathbb{R}^k)$. Suppose there is $r_0>0$ such that $p_u(\boldsymbol{u})>0$ for all $\|\boldsymbol{u}\|<r_0$. Then, for every $R<\infty$, there exists $m_R\in\mathbb{N}$ such that the $m_R$-fold convolution density $p_u^{m_R*}$ is continuous and strictly positive on $\overline B(0,R)$. Consequently,
\[
    \inf_{\|\boldsymbol{z}\|\le R}p_u^{m_R*}(\boldsymbol{z})>0.
\]
\end{lemma}

\begin{proof}
Since $p_u\in L^2(\mathbb{R}^k)$, the two-fold convolution $g=p_u^{2*}$ is continuous. Moreover, $g(\boldsymbol{z})>0$ for all $\|\boldsymbol{z}\|<r_0$: for such $\boldsymbol{z}$, the set $B(0,r_0)\cap\{\boldsymbol{u}:\boldsymbol{z}-\boldsymbol{u}\in B(0,r_0)\}$ has positive Lebesgue measure, and the convolution integrand is strictly positive on this set. Choose $0<r<r_0$. If $L$ is a positive integer and $\|\boldsymbol{z}\|<Lr$, then $\boldsymbol{z}$ can be written as a sum of $L$ vectors in $B(0,r)$; integrating over small neighborhoods of those vectors shows that $g^{L*}(\boldsymbol{z})>0$. Taking $L$ with $R<Lr$ gives a continuous convolution density $g^{L*}=p_u^{(2L)*}$ that is positive on the compact set $\overline B(0,R)$, and hence has a positive minimum there.
\end{proof}

\begin{lemma}[Small Set Condition for Continuous Covariate]
    \label{lemma_small_set_condition_continuous_covariate}

Under the strict feasibility condition in Proposition~\ref{prop:ergodicity} and Assumptions~\ref{ass:covariates} and \ref{ass:continuous}, for any bounded set $C \subset \mathbb{R}^k$, there exist a constant $n_0 \in \mathbb{N}$, a constant $\epsilon>0$, and a probability measure $\nu(\cdot)$ such that
    \begin{equation*}
        P^{n_0}(\boldsymbol{d},A) \geq \epsilon \nu(A) ,
    \end{equation*}
for any $\boldsymbol{d} \in C$ and any measurable set $A \subset \mathbb{R}^k$. Consequently, any bounded set $C$ is a small set.
\end{lemma}

\begin{proof}
Let $\mathbb P_{\boldsymbol d}$ denote the law of the imbalance process started at $\boldsymbol D_0=\boldsymbol d$. Since $\phi(\boldsymbol d,\boldsymbol t)\geq\iota>0$, we have
\begin{align*}
P^n(\boldsymbol d,A)
&=\mathbb P_{\boldsymbol d}(\boldsymbol D_n\in A)\\
&\geq \mathbb P_{\boldsymbol d}\left(
Z_1=\cdots=Z_n=1,\,
\boldsymbol d+
\sum_{i=1}^{n}(\boldsymbol x_i-\boldsymbol\mu_{EC})
\in A
\right).
\end{align*}
Denote $p_u(\boldsymbol{t}) = p(\boldsymbol{t}+\boldsymbol{\mu}_{EC})$. By Assumption~\ref{ass:continuous}, $p_u\in L^2(\mathbb{R}^k)$ and is strictly positive on an open neighborhood of the origin. By the boundedness of the set $C$, there exists some constant $M>0$ such that $\|\boldsymbol{d}\| \leq M$ for all $\boldsymbol{d} \in C$. Apply Lemma~\ref{lemma:convolution-support} with $R=M+1$. Then there exists $n_0\in\mathbb{N}$ such that $p_u^{n_0*}$ is continuous and strictly positive on $\overline{B(0,M+1)}$. Let
    \begin{equation*}
        c^* = \min_{z \in \overline{B(0, M+1)}} p_u^{n_0 *}(z) > 0 .
    \end{equation*}
Substituting this bound gives, for any $\boldsymbol{d} \in C \subset \overline{B(0,M)}$ and any measurable set $A \subset \mathbb{R}^k$,
    \begin{align*}
        & \quad P^{n_0}(\boldsymbol{d}, A)
        \geq \iota^{n_0} P \left( \sum_{i=1}^{n_0} (\boldsymbol{x}_i - \boldsymbol{\mu}_{EC}) \in A - \boldsymbol{d} \right)
        = \iota^{n_0} \int_{A} p_u^{n_0 *}(\boldsymbol{y} - \boldsymbol{d}) d\boldsymbol{y} \\
        & \geq \iota^{n_0} \int_{A \cap B(0, 1)} p_u^{n_0 *}(\boldsymbol{y} - \boldsymbol{d}) d\boldsymbol{y}
        \geq \iota^{n_0} c^* \text{Vol}(A \cap B(0, 1))
        = \left[ \iota^{n_0} c^* \text{Vol}(B(0, 1)) \right] \nu(A) ,
    \end{align*}
where $\nu(A) = \text{Vol}(A \cap B(0, 1)) / \text{Vol}(B(0, 1))$ is a probability measure restricted to the unit ball $B(0, 1)$. Setting $\epsilon = \iota^{n_0} c^* \text{Vol}(B(0, 1))$, the desired condition holds when the covariate distribution is continuous.
\end{proof}

When Assumption~\ref{ass:discrete} holds, the state space is countable and defined as follows, with each $n_i$ a nonnegative integer:
\begin{equation*}
    S = \left\{ \boldsymbol{d} = \sum_{i=1}^{s} n_i (\boldsymbol{t}_i - \boldsymbol{\mu}_{EC}) \mid n_i \in \mathbb{N} \right\} .
\end{equation*}

\begin{lemma}[Petite Set Condition for Discrete Covariates]
    \label{lemma_small_set_condition_discrete_covariate}

Under the strict feasibility condition in Proposition~\ref{prop:ergodicity} and Assumptions~\ref{ass:covariates} and \ref{ass:discrete}, for any $\boldsymbol{d} \in S$, there exists a constant $n_0(\boldsymbol{d}) \in \mathbb{N}$ such that
    \begin{equation*}
        P^{n_0(\boldsymbol{d})}(\boldsymbol{d},0) > 0 .
    \end{equation*}
Moreover, any bounded set $C \subset S$ is finite. Consequently, any bounded set $C$ is a petite set.
\end{lemma}

\begin{proof}
Let the matrix $A = (\boldsymbol{t}_1 - \boldsymbol{\mu}_{EC}, \dots, \boldsymbol{t}_s - \boldsymbol{\mu}_{EC})$. Then the linear equation $A p = 0$ is solvable, and its solution can be expressed in the form
    \begin{equation*}
        p = \sum_{j=1}^{s-r} c_j v_j ,
    \end{equation*}
where $\{v_1, \dots, v_{s-r}\}$ is a basis of the null space of $A$, $r$ is the rank of $A$, and $c_j \in \mathbb{R}$ are some coefficients. Because the entries of the matrix $A$ are rational, we can choose the basis vectors $\{v_1, \dots, v_{s-r}\}$ to have rational entries as well.

By Assumption~\ref{ass:covariates}, $0$ lies in the interior of the convex hull of the columns of $A$. Since the set
    \begin{equation*}
        \left\{ A p \mid p = (p_1, \dots, p_s)^T, p_i > 0, \sum_{i=1}^{s} p_i = 1 \right\}
    \end{equation*}
is convex and its closure equals the convex hull of the columns of $A$, it follows that this set contains the interior of the convex hull of the columns of $A$. Thus, there exists some $p^* = (p_1^*, \dots, p_s^*)^T \in \mathbb{R}^s$ with $p_i^* > 0$ and $\sum_{i=1}^{s} p_i^* = 1$, it holds that
    \begin{equation*}
        0 = A p^* .
    \end{equation*}

Thus, $p^* = \sum_{j=1}^{s-r} c_j^* v_j$ holds for some coefficients $c_j^* \in \mathbb{R}$. By perturbing the coefficients $c_j^*$ slightly, we can obtain rational coefficients $c_j^{**} \in \mathbb{Q}$ such that the vector $p^{**} = \sum_{j=1}^{s-r} c_j^{**} v_j$ has strictly positive and rational entries. The form $p^{**} = \sum_{j=1}^{s-r} c_j^{**} v_j$ implies that $p = p^{**}$ is also a solution to the equation $A p = 0$.

By multiplying by a suitable positive integer $m$, we may assume that all $m p_i^{**}$ are positive integers. Let $k_i = m p_i^{**}$ for $i=1, \dots, s$. By construction, each $k_i$ is a positive integer and satisfies
    \begin{equation*}
        \sum_{i=1}^{s} k_i (\boldsymbol{t}_i - \boldsymbol{\mu}_{EC}) = m A p^{**} = 0 .
    \end{equation*}
For any $\boldsymbol{d} \in S$, there exist non-negative integers $\{ n_i \}$ such that
    \begin{equation*}
        \boldsymbol{d} = \sum_{i=1}^{s} n_i (\boldsymbol{t}_i - \boldsymbol{\mu}_{EC}) .
    \end{equation*}
Then there exists some integer $n_d$ such that $n_d > \max_{1 \leq i \leq s} \frac{n_i}{k_i}$. Thus, we have
    \begin{equation*}
        \boldsymbol{d} + \sum_{i=1}^{s} (k_i n_d - n_i) (\boldsymbol{t}_i - \boldsymbol{\mu}_{EC}) = 0 .
    \end{equation*}

Let $n_0(\boldsymbol{d}) = n_d \sum_{i=1}^{s} k_i - \sum_{i=1}^{s} n_i \in \mathbb{N}$. Then, by the same argument as in the proof of Lemma \ref{lemma_small_set_condition_continuous_covariate}, we can show that
    \begin{align*}
        P^{n_0(\boldsymbol{d})}(\boldsymbol{d}, 0)
        & \geq \iota^{n_0(\boldsymbol{d})} P \left( \boldsymbol{d} + \sum_{i=1}^{n_0(\boldsymbol{d})} (\boldsymbol{x}_i - \boldsymbol{\mu}_{EC}) = 0 \right) \\
        & \geq \iota^{n_0(\boldsymbol{d})} P \left( \bigcap_{j=1}^{s} \left\{ \text{exactly } (k_j n_d - n_j) \text{ of the } \boldsymbol{x}_i \text{ equal } \boldsymbol{t}_j \right\} \right) \\
        & \geq \iota^{n_0(\boldsymbol{d})} \prod_{i=1}^{s} \mathbb{P}(\boldsymbol{x} = \boldsymbol{t}_i)^{k_i n_d - n_i}
        > 0 .
    \end{align*}

For the finiteness of any bounded set $C \subset S$, note that there exists a common denominator $L \in \mathbb{Z}^+$ such that all step vectors $\boldsymbol{t}_i - \boldsymbol{\mu}_{EC}$ lie in the lattice $\frac{1}{L}\mathbb{Z}^k$. Consequently, $S \subset \frac{1}{L}\mathbb{Z}^k$. Since the intersection of any discrete lattice with a bounded set in Euclidean space is finite, any bounded subset $C \subset S$ must be finite.

Therefore, set the transition kernel
    \begin{equation*}
        K_{1/2}(\boldsymbol{d},A) = \sum_{n=0}^\infty \frac{P^n(\boldsymbol{d},A)}{2^{n+1}} .
    \end{equation*}
Then, for any bounded set $C \subset S$, we have
    \begin{equation*}
        \inf_{\boldsymbol{d} \in C} K_{1/2}(\boldsymbol{d},0)
        \geq \inf_{\boldsymbol{d} \in C} \frac{P^{n_0(\boldsymbol{d})}(\boldsymbol{d},0)}{2^{n_0(\boldsymbol{d})+1}}
        > 0 .
    \end{equation*}
In conclusion, the set $C$ is $\nu_{1/2}$-petite for the measure $\nu_{1/2}(\cdot) = \epsilon \delta_0(\cdot)$, where $\delta_0$ is the Dirac measure at state $0$ and the constant $\epsilon > 0$.
\end{proof}

The drift verification and the continuous/discrete small-set arguments establish the hypotheses of the Foster--Lyapunov drift-minorization criterion for $\psi$-irreducible, aperiodic Markov chains \citep{meyn2009markov}. We consider the chain on its reachable state space. The minorization construction gives $\psi$-irreducibility on that space. The lower and upper feasibility bounds $\iota\le \phi(\boldsymbol{d},\boldsymbol{x})\le 1-\iota$ imply that every reachable state has a positive probability of remaining fixed after the next candidate, which gives aperiodicity. The drift calculation gives
\[
    PV(\boldsymbol{d})-V(\boldsymbol{d})
    \le -\epsilon + b\mathbf{1}_C(\boldsymbol{d})
\]
for $V(\boldsymbol{d})=\|\boldsymbol{d}\|^2$ and a bounded petite set $C$; the continuous case has a small set, and the discrete case has a petite set. Therefore the chain is positive Harris recurrent and has a unique invariant distribution $\pi$. The drift calculation also has the reward form, after enlarging $C$ if needed,
\[
    PV(\boldsymbol{d})-V(\boldsymbol{d})
    \le -a\|\boldsymbol{d}\|+b\mathbf{1}_C(\boldsymbol{d})
\]
for constants $a,b>0$. The corresponding drift theorem gives
\[
    \int \|\boldsymbol{d}\|\,\pi(d\boldsymbol{d})<\infty.
\]
This proves the positive Harris recurrence, uniqueness, ergodicity, and finite-first-moment conclusion in Proposition~\ref{prop:ergodicity}.

\subsection{Proof of Theorem~\ref{thm:balance}}

\begin{proof}
We first establish linear growth of the enrolled sample size. Let
\[
  \mathcal F_i
  =\sigma(\boldsymbol x_k,Z_k:1\le k\le i).
\]
The strict feasibility condition gives $p_{\min}=p_{\mathrm{avg}}-\Delta p/2>0$. For each candidate, write
\[
  q_i
  =\Prob(Z_i=1\mid\mathcal F_{i-1},\boldsymbol x_i)
  =\phi(\boldsymbol D_{i-1},\boldsymbol x_i)
  \ge p_{\min},
  \qquad
  e_i=Z_i-q_i.
\]
Then $e_i$ is $\mathcal F_i$-measurable, $\E(e_i\mid\mathcal F_{i-1})=0$ and $|e_i|\le1$. Moreover,
\[
  R_N
  =\sum_{i=1}^N q_i+\sum_{i=1}^N e_i,
  \qquad
  \sum_{i=1}^N q_i\ge p_{\min}N.
\]
For any $0<c<p_{\min}$, Azuma--Hoeffding's inequality gives
\[
  \Prob(R_N<cN)
  \le
  \Prob\left\{
  \sum_{i=1}^N e_i\le-(p_{\min}-c)N
  \right\}
  \le \exp(-C_cN)
\]
for some $C_c>0$. Thus $R_N\ge cN$ with probability tending to one. By the Borel--Cantelli lemma, the same bound holds eventually almost surely.

For part (i), Proposition~\ref{prop:ergodicity} establishes that the imbalance chain is positive Harris recurrent and aperiodic on its reachable state space, with a unique invariant distribution $\pi$. Hence $\boldsymbol D_N$ converges in distribution to $\pi$. The sequence is therefore tight, so $\|\boldsymbol D_N\|=O_p(1)$. Since
\[
  \bar{\boldsymbol x}_{SAT}^{(N)}-\boldsymbol\mu_{EC}
  =R_N^{-1}\boldsymbol D_N
\]
and $R_N\ge cN$ with probability tending to one,
\[
  \bar{\boldsymbol x}_{SAT}^{(N)}-\boldsymbol\mu_{EC}
  =O_p(N^{-1}).
\]

For part (ii), bounded covariates imply that the one-step increments of $\boldsymbol D_n$ are uniformly bounded. Let
\[
  \Delta_n
  =\|\boldsymbol D_{n+1}\|-\|\boldsymbol D_n\|,
\]
and choose $B_\Delta<\infty$ such that $|\Delta_n|\le B_\Delta$. Appendix~\ref{app:drift} gives
\[
  \E\left[
  \|\boldsymbol D_{n+1}\|^2-\|\boldsymbol D_n\|^2
  \mid\boldsymbol D_n
  \right]
  \le-\kappa\|\boldsymbol D_n\|+b_0,
\]
where
\[
  \kappa
  =\Delta p\,c_*
   -2p_{\mathrm{avg}}\|\boldsymbol\delta\|>0,
  \qquad
  b_0
  =\left(p_{\mathrm{avg}}+\frac{\Delta p}{2}\right)
   \E\|\boldsymbol u\|^2<\infty.
\]
Using the identity
\[
  \|\boldsymbol D_{n+1}\|^2-\|\boldsymbol D_n\|^2
  =
  2\|\boldsymbol D_n\|\Delta_n+\Delta_n^2
\]
and $\Delta_n^2\ge0$, we obtain, whenever $\|\boldsymbol D_n\|>0$,
\[
  \E(\Delta_n\mid\boldsymbol D_n)
  \le
  -\frac{\kappa}{2}
  +\frac{b_0}{2\|\boldsymbol D_n\|}.
\]
Choose $K>2b_0/\kappa$ and set $\gamma=\kappa/4$. Then
\[
  \E(\Delta_n\mid\boldsymbol D_n)\le-\gamma
  \qquad
  \text{whenever }\|\boldsymbol D_n\|>K.
\]

The bounded-increment condition gives the uniform Taylor bound
\[
  \E(e^{\alpha\Delta_n}\mid\boldsymbol D_n)
  \le
  1+\alpha\E(\Delta_n\mid\boldsymbol D_n)
  +C_{B_\Delta}\alpha^2
\]
for all sufficiently small $\alpha>0$, where $C_{B_\Delta}<\infty$ depends only on $B_\Delta$. Choose $\alpha$ small enough that $C_{B_\Delta}\alpha^2\le\alpha\gamma/2$ and $\eta=\alpha\gamma/2\in(0,1)$. Outside the set $\{\|\boldsymbol d\|\le K\}$,
\[
  \E(e^{\alpha\Delta_n}\mid\boldsymbol D_n)
  \le1-\eta.
\]
Thus the exponential Lyapunov function
\[
  W(\boldsymbol d)=\exp(\alpha\|\boldsymbol d\|)
\]
satisfies
\[
  \E\{W(\boldsymbol D_{n+1})\mid\boldsymbol D_n\}
  \le(1-\eta)W(\boldsymbol D_n)+b_1
\]
for some $b_1<\infty$. Indeed, within the bounded set, $W(\boldsymbol D_{n+1})\le \exp\{\alpha(K+B_\Delta)\}$, so $b_1$ can be chosen uniformly over all states.

Iterating the inequality gives
\[
  \E W(\boldsymbol D_n)
  \le
  (1-\eta)^nW(\boldsymbol D_0)+\frac{b_1}{\eta}.
\]
Since $\boldsymbol D_0=\boldsymbol0$, there is a constant $C_0<\infty$ such that
\[
  \sup_{n\ge0}
  \E\exp(\alpha\|\boldsymbol D_n\|)\le C_0.
\]
Markov's inequality therefore gives
\[
  \Prob(\|\boldsymbol D_n\|>x)
  \le C_0e^{-\alpha x},
  \qquad x\ge0.
\]
Taking a union bound over the first $N$ steps,
\[
  \Prob\left(
  \max_{1\le n\le N}\|\boldsymbol D_n\|>x
  \right)
  \le NC_0e^{-\alpha x}.
\]
For $N\ge2$, set $x=L\log N$ with $L>2/\alpha$. Then
\[
  \Prob\left(
  \max_{1\le n\le N}\|\boldsymbol D_n\|>L\log N
  \right)
  \le C_0N^{1-\alpha L}.
\]
The right side is summable in $N$. The Borel--Cantelli lemma yields $\|\boldsymbol D_N\|=O(\log N)$ almost surely. Combining this with the eventual almost-sure bound $R_N\ge cN$ gives
\[
  \bar{\boldsymbol x}_{SAT}^{(N)}-\boldsymbol\mu_{EC}
  =R_N^{-1}\boldsymbol D_N
  =O(N^{-1}\log N)
  \qquad\text{almost surely}.
\]
\end{proof}

\subsection{Centering of the survival score}
\label{app:tte-regularity}

We use the assumptions in the main text. For an enrolled subject, define
\[
  N_i(t)=1\{F_i\le t,F_i\le C_i\},\qquad
  Y_i^{\mathrm{risk}}(t)=1\{F_i\ge t,C_i\ge t\},
  \qquad 0\le t\le\tau.
\]
Since $C_i\le\tau$, the score contribution can be written as
\[
  M_i
  =\int_0^\tau
    \{dN_i(t)-Y_i^{\mathrm{risk}}(t)d\Lambda_0(t)\}
  =\Delta_i-\Lambda_0(Y_i),
\]
where $Y_i=\min(F_i,C_i)$ is the observed time.

Let $G_j(t)=\Prob(C_i\ge t\mid s_i=j)$. Independence of event and censoring times within strata gives
\[
  \E\{dN_i(t)\mid s_i=j\}
  =S_j(t)G_j(t)\lambda_j(t)\,dt,\qquad
  \E\{Y_i^{\mathrm{risk}}(t)\mid s_i=j\}
  =S_j(t)G_j(t).
\]
Consequently,
\[
  \mu_j
  =\int_0^\tau S_j(t)G_j(t)
    \{\lambda_j(t)-\lambda_0(t)\}\,dt.
\]

Under $H_0$, the internal survival function at the external stratum proportions equals $S_0$:
\[
  \sum_jp_j^*S_j(t)=S_0(t).
\]
Absolute continuity allows differentiation, yielding
\[
  \sum_jp_j^*S_j(t)\lambda_j(t)
  =S_0(t)\lambda_0(t)
\]
for almost every $t\in[0,\tau]$. Assumption~\ref{ass:finite-observation}(a) gives $G_j=G$, so
\begin{align*}
  \sum_jp_j^*\mu_j
  &=
  \int_0^\tau G(t)
  \left\{
    \sum_jp_j^*S_j(t)\lambda_j(t)
    -\lambda_0(t)\sum_jp_j^*S_j(t)
  \right\}dt\\
  &=
  \int_0^\tau G(t)
  \{S_0(t)\lambda_0(t)-\lambda_0(t)S_0(t)\}\,dt
  =0.
\end{align*}

Individual stratum means may be nonzero because each stratum is compared with the pooled benchmark. Under $H_0$, these means cancel when weighted by the external proportions. Exact calibration therefore gives
\[
  U=m\sum_jp_j^*\bar M_j
   =m\sum_jp_j^*(\bar M_j-\mu_j),
\]
so the score combines fluctuations around the stratum means using the fixed external proportions.

\subsection{Proof of Theorem~\ref{thm:tte}}

\begin{proof}
Let $\mathcal H=\sigma(s_k,\xi_k:k\ge1)$ be the $\sigma$-field generated by all candidate strata and the random variables used to make enrollment decisions. Under the candidate and enrollment model, the accepted indices, stopping index, stratum counts and weights are $\mathcal H$-measurable. Conditional on $\mathcal H$, the enrolled patients' score contributions are independent and each has the fixed conditional distribution of $M$ given its stratum, with mean $\mu_j$ and variance $v_j$.

Appendix~\ref{app:tte-regularity} gives $\sum_jp_j^*\mu_j=0$. Thus
\[
  U=\sum_j\frac{mp_j^*}{n_j}
        \sum_{i:s_i=j}(M_i-\mu_j),
  \qquad
  \E(U\mid\mathcal H)=0,
\]
and
\[
  V_m^*=\var(U\mid\mathcal H)
  =m^2\sum_j\frac{(p_j^*)^2v_j}{n_j},
  \qquad
  V_m^*/m\to_p V.
\]
The bound $|M_i|\le K=1+\Lambda_0(\tau)$ implies
\[
  \max_j\frac{2K\,mp_j^*/n_j}{\sqrt{V_m^*}}
  \le \max_j\frac{2K}{\sqrt{n_jv_j}}
  \ \xrightarrow{p}\ 0.
\]
Consequently the conditional Lindeberg condition holds in probability. More explicitly, every subsequence has a further subsequence on which $n_j/m\to q_j$ almost surely for every $j$. For almost every such design realization, the ordinary triangular-array Lindeberg theorem applies to the conditionally independent centered terms. The conditional normal approximation therefore converges in probability along the full sequence; boundedness of conditional probabilities permits integration. This yields $U/\sqrt{V_m^*}\to_d N(0,1)$ and hence $U/\sqrt m\to_d N(0,V)$.

For variance estimation, conditional independence and boundedness give
\[
  \var(\bar M_j\mid\mathcal H)\le K^2/n_j,
  \qquad
  \var\left(
  n_j^{-1}\sum_{i:s_i=j}M_i^2
  \,\middle|\,\mathcal H
  \right)\le K^4/n_j.
\]
On the event $n_j\ge2$, conditional Chebyshev bounds and $n_j\to_p\infty$ show that the first and second sample moments converge to the fixed stratum moments. Hence $s_j^2\to_p v_j$. Since $J$ is fixed,
\[
  \frac{\widehat V_{\mathrm{new}}}{m}
  =\sum_j\frac{(p_j^*)^2s_j^2}{n_j/m}
  \to_p V.
\]
Finally, Slutsky's theorem proves the studentized limit.
\end{proof}

\subsection{Proof of Proposition~\ref{prop:conventional-limit}}
\label{app:conventional-limit}

\begin{proof}
Use the $\sigma$-field $\mathcal H$ from the proof of Theorem~\ref{thm:tte}. Let $L=\Lambda_0(\tau)$ and let $\bar e_j$ be the average of $\Lambda_0(Y_i)$ over enrolled subjects in stratum $j$. Conditional on $\mathcal H$, these terms are independent with mean $e_j$ and lie in $[0,L]$. Therefore,
\[
  \E(\bar e_j\mid\mathcal H)=e_j,
  \qquad
  \var(\bar e_j\mid\mathcal H)\le \frac{L^2}{n_j}.
\]
Because $n_j/m\to_p q_j>0$, conditional Chebyshev bounds give $\bar e_j\to_p e_j$. Substituting $w_j=mp_j^*/n_j$ gives
\[
  \frac{\widehat V_{\mathrm{conv}}}{m}
  =\sum_j\frac{(p_j^*)^2}{n_j/m}\,\bar e_j
  \ \xrightarrow{p}\
  \sum_j\frac{(p_j^*)^2e_j}{q_j}.
\]
Subtracting $V=\sum_j(p_j^*)^2v_j/q_j$ proves Equation~\eqref{eq:conventional-gap}. This limit calculation does not require $H_0$.

For the second-moment identity, construct a single draw from the target population: sample stratum $s$ with probabilities $\boldsymbol p^*$, then draw $(F,C)$ from its conditional law. Write $\Prob_*$ and $\E_*$ for probability and expectation under this target population. Assumption~\ref{ass:finite-observation}(a) gives
\[
  \Prob_*(s=j,F>u,C\ge t)
  =p_j^*S_j(u)G(t).
\]
Thus $C$ is independent of $(s,F)$. Under $H_0$, the marginal survival function of $F$ is $S_0$. Define, for $0\le t\le\tau$,
\[
  N(t)=1\{F\le t,F\le C\},
  \qquad
  R(t)=1\{F\ge t,C\ge t\},
  \qquad
  A(t)=\int_0^t R(u)\,d\Lambda_0(u).
\]
In the observed event and censoring filtration that does not include $s$, the event intensity is $R(t)\lambda_0(t)$. Consequently, $\mathcal M(t)=N(t)-A(t)$ is a square-integrable martingale. The compensator is continuous, so the predictable quadratic variation of $\mathcal M$ is $\langle\mathcal M\rangle(t)=A(t)$. At the fixed horizon, $\mathcal M(\tau)=M$ and $A(\tau)=\Lambda_0(Y)$, where $Y=\min(F,C,\tau)$. The martingale second-moment identity yields
\[
  \E_*(M^2)
  =\E_*\{\langle\mathcal M\rangle(\tau)\}
  =\E_*\{\Lambda_0(Y)\}
  =\sum_jp_j^*e_j.
\]
Conditioning the left side on $s$ instead gives
\[
  \E_*(M^2)
  =\sum_jp_j^*(v_j+\mu_j^2).
\]
These equalities prove Equation~\eqref{eq:mixture-second-moment}. The martingale argument uses the target population with stratum unobserved, rather than the conditional distribution within a stratum.

For Remark~\ref{rem:conventional-conservative}, suppose $H_0$ holds and $q_j=p_j^*$ for every $j$. The two limits then reduce to $V=\sum_jp_j^*v_j$ and $V_{\mathrm{conv}}=\sum_jp_j^*e_j$, so
\[
  V_{\mathrm{conv}}-V
  =\sum_jp_j^*\mu_j^2\ge0.
\]
In particular, $V_{\mathrm{conv}}\ge V>0$. Theorem~\ref{thm:tte} and Slutsky's theorem give the stated normal limit. Continuity of the normal distribution at the two critical values gives the limiting rejection probability. Since all target proportions are positive, equality holds exactly when every $\mu_j$ is zero.
\end{proof}

\subsection{Proof of Theorem~\ref{thm:cont}}

\begin{proof}
For enrolled subjects,
\[
  Y_i(1)
  =\alpha_*+
  \boldsymbol{\beta}_*^\top
  (\boldsymbol{B}_i-\boldsymbol{\mu}_{EC})
  +\varepsilon_{*,i}.
\]
Under $H_0$, $\alpha_*=\bar Y_{EC}(0)$. Therefore,
\[
  \hat\theta_{\mathrm{cal}}-\bar Y_{EC}(0)
  =\bar\varepsilon_{*,SAT}
  +(\boldsymbol{\beta}_*-\hat{\boldsymbol{\beta}})^\top
   (\bar{\boldsymbol{B}}_{SAT}-\boldsymbol{\mu}_{EC}),
\]
where $\bar\varepsilon_{*,SAT} =R_N^{-1}\sum_{i\in\mathcal I_N}\varepsilon_{*,i}$.

We first establish consistency of the OLS coefficients. Write $\boldsymbol H_i=(1,\boldsymbol B_i^\top)^\top$ and define
\[
  Q_N
  =R_N^{-1}\sum_{i\in\mathcal I_N}
    \boldsymbol H_i\boldsymbol H_i^\top,
  \qquad
  \boldsymbol c_N
  =R_N^{-1}\sum_{i\in\mathcal I_N}
    \boldsymbol H_i\varepsilon_{*,i}.
\]
Lemma~\ref{lemma:stationary-enrolled-limits} gives $Q_N\to_p Q$, where $Q$ is positive definite. Using the filtration $\mathcal F_i$ in that lemma, Assumption~\ref{ass:cont-reg} implies
\[
  \E\{
  Z_i\boldsymbol H_i\varepsilon_{*,i}
  \mid\mathcal F_{i-1}
  \}=\boldsymbol 0.
\]
Moreover, $\E\|\boldsymbol H_i\varepsilon_{*,i}\|<\infty$ by Cauchy--Schwarz inequality and the second-moment conditions. Truncation gives a martingale law of large numbers: the bounded centered terms have vanishing average variance, and the expected tail contributions vanish uniformly in $i$ because the candidate records are i.i.d. Thus
\[
  N^{-1}\sum_{i=1}^N
  Z_i\boldsymbol H_i\varepsilon_{*,i}
  \to_p\boldsymbol 0.
\]
Since $R_N/N\to_p\rho_\pi>0$, it follows that $\boldsymbol c_N\to_p\boldsymbol 0$. The OLS normal equations then imply $\hat{\boldsymbol\beta}-\boldsymbol\beta_*=o_p(1)$.

By Theorem~\ref{thm:balance} and the enrollment-rate limit,
\[
  \bar{\boldsymbol B}_{SAT}-\boldsymbol\mu_{EC}
  =R_N^{-1}\boldsymbol D_N
  =O_p(N^{-1}).
\]
Consequently,
\[
  \sqrt{R_N}
  (\boldsymbol\beta_*-\hat{\boldsymbol\beta})^\top
  (\bar{\boldsymbol B}_{SAT}-\boldsymbol\mu_{EC})
  =o_p(1).
\]
The leading term is therefore the enrolled residual average. Lemma~\ref{lemma:stationary-enrolled-limits} gives
\[
  \sqrt{R_N}\bar\varepsilon_{*,SAT}
  \xrightarrow{d}
  \mathcal N(0,\sigma_{\mathrm{cal}}^2).
\]
Combining these results yields
\[
  \sqrt{R_N}
  \left(\hat\theta_{\mathrm{cal}}-\bar Y_{EC}(0)\right)
  \xrightarrow{d}
  \mathcal N(0,\sigma_{\mathrm{cal}}^2).
\]

For variance estimation, let $\hat\varepsilon_i$ be the residuals from the enrolled-sample OLS regression of $Y$ on $1$ and $\boldsymbol B$. On the event that $Q_N$ is invertible, whose probability tends to one, the OLS residual identity gives
\[
  R_N^{-1}\sum_{i\in\mathcal I_N}\hat\varepsilon_i^2
  =
  R_N^{-1}\sum_{i\in\mathcal I_N}\varepsilon_{*,i}^2
  -\boldsymbol c_N^\top Q_N^{-1}\boldsymbol c_N.
\]
The first term converges to $\sigma_{\mathrm{cal}}^2$ by Lemma~\ref{lemma:stationary-enrolled-limits}, and the second converges to zero because $\boldsymbol c_N\to_p\boldsymbol 0$ and $Q_N\to_p Q$. Since the feature dimension $q$ is fixed and $R_N\to_p\infty$,
\[
  \hat\sigma_{\mathrm{cal}}^2
  =
  \frac{1}{R_N-q-1}
  \sum_{i\in\mathcal I_N}\hat\varepsilon_i^2
  \xrightarrow{p}
  \sigma_{\mathrm{cal}}^2.
\]
Finally, $\sigma_{\mathrm{cal}}^2>0$ and Slutsky's theorem give
\[
  \frac{
  \sqrt{R_N}
  \left(\hat\theta_{\mathrm{cal}}-\bar Y_{EC}(0)\right)
  }{\hat\sigma_{\mathrm{cal}}}
  \xrightarrow{d}\mathcal N(0,1).
\]
\end{proof}

\end{document}

%% file: case_current_table.tex
\begin{tabular}{lrrrr}
\toprule
Design/analysis & $H_0$ rejection & $H_1$ rejection & $H_0$ SD & ESS \\
\midrule
Unadjusted & 0.430 & 0.519 & 1.142 & 100.0 \\
MAIC-JK & 0.056 & 0.875 & 1.034 & 83.4 \\
CAE-conventional & 0.031 & 0.813 & 0.910 & 97.4 \\
CAE-new & 0.052 & 0.907 & 1.022 & 97.4 \\
CAE-JK & 0.054 & 0.913 & 1.024 & 97.4 \\
\bottomrule
\end{tabular}